\documentclass[11pt]{iopart}

\usepackage{bm}
\usepackage{amssymb}
\usepackage{amsopn}
\usepackage{hyperref}
\usepackage{feynmp}
\usepackage{ifpdf}

\ifpdf
\else 
\fi

\renewcommand{\etal}{\emph{et al.}}

\newcommand{\Mp}{M_{\mathrm{P}}}

\newcommand{\LambdaPhys}{\Lambda_{\mathrm{P}}}

\newcommand{\meff}{M_{\mathrm{eff}}}
\newcommand{\Niso}{N_{\mathrm{s}}}

\renewcommand{\d}{\mathrm{d}}
\newcommand{\vect}[1]{\bm{\mathrm{{#1}}}}
\renewcommand{\e}[1]{\mathrm{e}^{{#1}}}
\newcommand{\im}{\mathrm{i}}

\renewcommand{\geq}{\geqslant}

\makeatletter
\newcommand\numberwithin[2]{\@addtoreset{#1}{#2}}
\makeatother
\numberwithin{footnote}{subsection}

\newcommand{\parag}[1]{\paragraph{\upshape\textsf{{#1}}}}

\begin{document}
	\title{Infrared effects in inflationary correlation functions}
	\author{David Seery}
	\address{Astronomy Centre, University of Sussex, Brighton,
	United Kingdom BN1 9QH}
	\eads{\mailto{D.Seery@sussex.ac.uk}}
	\begin{abstract}
		In this article, I briefly review the status of infrared
		effects which occur when using inflationary models to
		calculate initial conditions for a subsequent hot, dense
		plasma phase.
		Three types of divergence have been identified in the
		literature: secular, ``time-dependent'' logarithms, which grow with
		time spent outside the horizon; ``box-cutoff'' logarithms,
		which encode a dependence on the infrared cutoff when
		calculating in a finite-size box; and ``quantum''
		logarithms,
		which depend on the ratio of a scale characterizing new
		physics to the scale of whatever process is under consideration,
		and
		whose interpretation is the same as conventional
		field theory.
		I review the calculations in which these divergences appear,
		and discuss the methods which have been developed to deal with them.

		\vspace{3mm}
		\begin{flushleft}
			\textsf{\textbf{Keywords}:
			Inflation,
			Cosmological perturbation theory,
			Physics of the early universe,
			Quantum field theory in curved spacetime.}
		\end{flushleft}
	\end{abstract}
	\maketitle
	
	\begin{fmffile}{diags}
	
	\setcounter{footnote}{0}
	\section{Introduction}
	\label{sec:intro}
	
	Looking out at the universe from the vantage point of Earth,
	we see a small fluctuation $\delta T$ in the temperature
	of the cosmic microwave background radiation (CMB). 
	Over the last decade, advances in observational astronomy have allowed
	us to commence a detailed study
	of $\delta T$'s statistical properties.
	We aim to compare these properties with those of small
	fluctuations which, according to our present ideas, are expected to
	have existed in the very early universe.
	
	These fluctuations evolve in a hot, dense
	plasma of tightly coupled baryons and photons. If we suppose the
	plasma era was preceded by an epoch of primordial inflation, then
	initial conditions for these fluctuations can be calculated from
	the parameters of the inflationary model.
	The variance predicted by this approach
	was computed in a now-classic series of early papers
	\cite{Bardeen:1983qw,Guth:1982ec,Hawking:1982cz,Hawking:1982my,
	Mukhanov:1985rz,Sasaki:1986hm},
	and later extended to the skewness
	\cite{Allen:1987vq,Falk:1992sf,Gangui:1993tt,Maldacena:2002vr,
	Seery:2005gb,Chen:2006nt}
	and kurtosis
	\cite{Seery:2006vu,Arroja:2008ga,Seery:2008ax,Chen:2009bc}.
	With sufficiently precise experimental data in hand,
	we can work backwards from observations
	of these statistical properties to determine what the
	initial conditions must have been.
	The result is
	a realistic prospect of constraining
	parameters in certain inflationary models.
	
	This ambitious programme was begun in earnest following
	the arrival of high-quality CMB maps made by the
	\emph{Wilkinson Microwave Anisotropy Probe} (WMAP)
	\cite{Komatsu:2010fb}.
	Even more precise data
	are expected from the \emph{Planck Surveyor} satellite.
	Once it had become
	clear that the statistical properties of $\delta T$
	carried important information
	about whatever physics was operative during inflation,
	theorists were soon tempted to refine their calculations
	in the hope that additional detail could be extracted.
	A similar programme in electroweak physics had provided important
	clues about the unknown details of electroweak symmetry breaking,%
		\footnote{More details of this ``precision electroweak
		programme'' can be found
		in the review \emph{Electroweak model and constraints on new
		physics,} prepared by the Particle Data Group
		\cite{Amsler:2008zzb}.}
	and it was reasonable to explore the possibility
	of similar gains in cosmology.
	The observable statistical properties of the inflationary density
	perturbation are encoded in its $n$-point correlation functions,
	with the leading contribution to each function typically suppressed
	by $(n-1)$ powers of the small quantity $(H/\Mp)^2$ \cite{Jarnhus:2007ia},
	where $H$ is the Hubble rate during inflation and $\Mp
	= (8\pi G)^{-1/2}$ is the Planck mass.
	Two types of refinement were possible: either to calculate
	to lowest order in $(H/\Mp)^2$
	at progressively larger $n$, or to
	calculate higher order corrections with $n$ fixed.

	It had been known for a long time that perturbation theory beyond
	leading order in $(H/\Mp)^2$
	was complicated by troublesome infrared behaviour.
	Sasaki, Suzuki, Yamamoto \& Yokoyama considered a $\lambda \phi^n$
	theory with $n \geq 3$, coupled non-minimally to gravity,
	and evaluated each $n$-point function perturbatively
	\cite{Sasaki:1992ux,Suzuki:1992gi}.
	Sasaki {\etal} noticed that the $n$-point functions
	evaluated in de Sitter space had
	a potential divergence in the far future, which they called
	``superexpansionary.''
	We will rephrase the analysis of Sasaki {\etal} in modern
	notation and discuss the meaning of these divergences in
	\S\ref{sec:zoo-time} and \S\ref{sec:time}.

	Later, Mukhanov, Abramo \& Brandenberger studied a different
	sort of infrared effect---%
	the accumulation of long wavelength fluctuations in
	an expanding universe
	\cite{Mukhanov:1996ak,Abramo:1997hu}.
	(See also Abramo \& Woodard, Ref.~\cite{Abramo:1998hi}.)
	This differed from the analysis of Sasaki {\etal},
	which had been based on a purely geometrical effect.
	Whether such
	divergences are physical
	is complicated by the question of gauge,
	initially studied by
	Unruh \cite{Unruh:1998ic}.
	Later, Losic \& Unruh
	were able to give a gauge-invariant
	argument that infrared terms generically
	become large in spacetimes close to de Sitter
	\cite{Losic:2005vg,Losic:2006ht,Losic:2008ht,Unruh:2008zza}.
	We will discuss corrections of this type in
	\S\ref{sec:zoo-box} and \S\ref{sec:space}.
	Divergences of a third kind had long been studied by
	Prokopec,
	Tsamis, Woodard and their collaborators.
	These authors took the quantum nature of correlation functions
	seriously, and calculated loop corrections just as one would
	when studying scattering experiments.
	Unfortunately, the complexity of
	both the rules for calculating these loops
	and the computations themselves are somewhat
	greater than in Minkowski space.
	These calculations have been reported in a large literature.
	Together with quantum effects studied by others,
	including Boyanovsky, de Vega \& Sanchez,
	a recent sample can be found in
	Refs.~\cite{Prokopec:2002uw,Onemli:2002hr,Prokopec:2003bx,Onemli:2004mb,
	Brunier:2004sb,
	Boyanovsky:2005sh,Boyanovsky:2005px,Kahya:2005kj,
	Kahya:2006hc,Bilandzic:2007nb,
	Kahya:2006ui,Prokopec:2006ue}.
	We will consider effects of this type in \S\ref{sec:zoo-new}.

	Perturbative calculations of correlation functions
	from inflation had been studied in the 1980s by
	Allen, Grinstein \& Wise
	\cite{Allen:1987vq}
	and later by Falk, Rangarajan \& Srednicki
	\cite{Falk:1992sf}.
	The sophistication of such calculations increased
	dramatically
	in the years following Maldacena's calculation of the three-point
	function in single-field slow-roll inflation \cite{Maldacena:2002vr}.
	At least in part, this increase was a consequence of the more
	complicated models of inflation which had been developed during
	the 1990s and early 2000s, and which theorists hoped would
	yield a distinctive pattern of correlations
	\cite{Seery:2005wm,Chen:2006nt,Huang:2006eha,Chen:2009bc}.
	These developments are reviewed pedagogically in a recent
	article by Chen \cite{Chen:2010xk}.
	
	Since 2002, significant theoretical effort has been expended
	in refining our understanding of both inflationary correlation functions
	themselves, and the models which can generate significantly
	non-gaussian statistics.
	The calculations reviewed in this paper were a
	consequence of the reinvigoration of interest in infrared issues
	which followed this effort.
	Logarithms exhibiting secular growth with time were discovered
	in calculations of $n$-point correlation functions
	with $n \geq 3$
	\cite{Falk:1992sf,Zaldarriaga:2003my,
	Sloth:2006az,Sloth:2006nu,Seery:2008qj}.
	Other calculations, if taken at face value, gave divergent answers
	in the infinite-volume limit. Divergences could be avoided by carrying
	out the calculation in a finite box, but any such procedure left behind
	residual logarithms involving the box cutoff
	\cite{Boubekeur:2005fj,Lyth:2005fi,
	Sloth:2006az,Sloth:2006nu,Byrnes:2007tm,Lyth:2007jh,Seery:2007wf}.
	Further logarithms were encountered while studying the influence of
	unknown physics at energy scales even larger than the inflationary
	scale
	\cite{Weinberg:2005vy,Weinberg:2006ac,Chaicherdsakul:2006ui,Seery:2007we,
	Adshead:2008gk,Dimastrogiovanni:2008af,Seery:2008ms,Senatore:2009cf}.
	
	Infrared effects are not a phenomenon unique to de Sitter
	calculations, but are present in many applications of quantum field
	theory. Their appearance typically signals the
	presence of a non-trivial
	background, which cannot be described by asymptotic
	\emph{in}- and \emph{out}-states
	containing a definite number of particles.
	In scattering calculations, infrared divergences allow
	putative fixed-particle states
	to develop a slowly varying field, made out of an accumulation
	of arbitrarily many soft particles radiated on approach to or
	recoil from the scattering event.
	This ``Bloch--Nordsieck'' or ``initial state radiation''
	phenomenon is studied in many textbooks on
	quantum field theory \cite{Weinberg:1995mt}.
	A similar effect can occur in the confined phase of
	quantum chromodynamics, where soft particles can be radiated by
	partons moving inside nuclei. These particles are trapped by the
	strong QCD force, and change the background colour field.
	This effect may dramatically alter the mix of partons
	observed by an impinging probe such as a photon, and cannot be
	neglected in an accurate comparison with experiment
	\cite{Gribov:1972ri,Gribov:1972rt,Altarelli:1977zs,Kogut:1974ni,
	Dokshitzer:1977sg,Collins:1988wj}.
	The effective $W$ approximation is a related example, in which
	the same infrared effects control the $W^\pm$
	content of colliding nuclei. The results are very practical.
	For example, this method can be used to predict the rate
	of Higgs production from $W^+ W^-$ fusion \cite{Gunion:1989we}.
	
	Throughout this article, Planck's constant $\hbar$ and
	the speed of light $c$ are set to unity.
	Many expressions are written in terms of the reduced Planck
	mass, $\Mp^{-2} = 8 \pi G$.
	For brevity, a scalar field which contributes to the
	energy--momentum tensor $T_{ab}$ is referred to as `active,'
	whereas a field making no contribution to $T_{ab}$
	is referred to as a `spectator.' In a cosmological context,
	such scalars are isocurvature modes.

	\section{A zoology of logarithms}
	
	In this section, I briefly review the sources of logarithmic
	divergences in inflationary correlation functions.
	
	\subsection{Time-dependent logarithms}
	\label{sec:zoo-time}
	
	A decade after calculations of the 2-point function had determined
	the variance in $\delta T$ expected from an inflationary initial
	condition, the corresponding skewness was calculated
	by Falk, Rangarajan \& Srednicki \cite{Falk:1992sf}
	in an approximation where gravitational interactions were neglected.%
		\footnote{It was later argued by Gangui {\etal} \cite{Gangui:1993tt}
		and in more detail by Maldacena \cite{Maldacena:2002vr} that
		gravitational interactions in fact provide a dominant
		contribution to the skewness, but the contribution calculated
		by Falk {\etal} is still present, if subdominant.
		See Ref.~\cite{Chen:2010xk} or
		the review of inflationary non-gaussianities
		by Koyama in this volume \cite{Koyama:2010xj}.}
	At that time, data from the COBE satellite \cite{Smoot:1992td}
	had established the existence of temperature fluctuations in
	the microwave background at the level $\delta T \sim 10^{-5} T$,
	where $T \approx 2.75 \, \mathrm{K}$ was the average microwave temperature
	over the sky. In inflationary models $\delta T/T$ is closely
	related to the parameter $(H/\Mp)^2$.
	Since the skewness is proportional to $(H/\Mp)^4$,
	it was already evident that any non-gaussianity would be very small.
	
	Let us suppose inflation is driven by a single scalar field with
	potential $V$.
	We define the slow-roll parameters by
	\begin{equation}
		\epsilon \equiv \frac{\Mp^2}{2} \left( \frac{V'}{V} \right)^2 ,
		\quad
		\eta \equiv \Mp^2 \frac{V''}{V} , \quad \mbox{and}
		\quad
		\xi^2 \equiv \Mp^4 \frac{V' V'''}{V^2} ,
	\end{equation}
	and use a conformal time variable, $\tau$, related to cosmic time $t$
	by $\tau = \int^t_\infty \d t' / a(t')$.
	When measured in this variable,
	horizon crossing occurs for a mode of wavelength $k$ when
	$k \tau = -1$, and
	after $N$ e-folds outside the horizon we have $k \tau = - \e{-N}$.
	The infinite future of de Sitter space
	corresponds to $|k \tau| \rightarrow 0$.
	In this notation the result of
	Falk, Rangarajan \& Srednicki can be written
	\begin{equation}
		\fl
		\langle \delta \phi(\vect{k}_1) \delta\phi(\vect{k}_2)
		\delta \phi(\vect{k}_3) \rangle_\ast
		=
		(2\pi)^3 \delta(\vect{k}_1 + \vect{k}_2 + \vect{k}_3)
		\frac{1}{4\sqrt{2}}
		\frac{H_\ast^4}{\Mp}
		\frac{\xi^2}{\sqrt{\epsilon}} \frac{\sum_i k_i^3}{\prod_j k_j^3}
		\ln |k \tau_\ast| + \cdots ,
		\label{eq:falklog}
	\end{equation}
	where a subscript `$\ast$' denotes evaluation at the time
	$\tau_\ast$, and
	`$\cdots$' denotes terms of lower order in the slow-roll
	expansion, together with other terms which do not grow as
	$|k \tau_\ast| \rightarrow 0$
	\cite{Zaldarriaga:2003my,Seery:2008qj}.%
		\footnote{If our purpose were only
		to calculate the density fluctuation
		produced by a single-field model of inflation, then
		the logarithmic divergence in Eq.~\eref{eq:falklog} would be
		entirely spurious
		\cite{Unruh:1998ic}. The density fluctuation is determined by
		the comoving curvature perturbation, $\zeta$, which is conserved
		on superhorizon scales, $\dot{\zeta} = 0$, in the absence of
		entropy modes.
		Occasionally it has been argued that an estimate of the
		non-gaussian yield at the end of inflation can be obtained
		by setting $\ln|k \tau_\ast| \sim 60$, but
		this example shows clearly that unless care is taken
		to account for possible cancellations any such procedure can be
		quite misleading. The possibility of such cancellations
		when reasoning with cutoff-dependent terms in
		a general effective field theory
		was pointed out by Burgess \& London \cite{Burgess:1992gx}.
		In the inflationary context, the pitfalls of this
		procedure were noted in
		Ref.~\cite{Seery:2008qj}. I would like to thank Xingang Chen
		for discussions on this point.}
	Higher powers of $\ln |k\tau_\ast|$ may be generated at higher order
	in the slow-roll expansion, either by retaining such terms explicitly
	in the tree-level calculation or by including loop corrections,
	as we shall shortly explain.
	A similar effect can be observed in the massless limit of
	the calculation by Chen \& Wang
	\cite{Chen:2009zp}.
	
	The result is a power series in $\ln |k\tau_\ast|$
	with an apparent divergence in the limit $\tau_\ast \rightarrow 0$,
	as Sasaki {\etal} had predicted \cite{Sasaki:1992ux}.
	Series representations of this type 
	were later applied
	to inflationary correlation functions by Gong \& Stewart
	\cite{Gong:2001he,Gong:2002cx}, who obtained them by systematically
	solving Mukhanov's equation \cite{Mukhanov:1985rz}
	using a Green's function approach. Gong \& Stewart remarked that
	the correction linear in logarithms was typically proportional to
	$\sim \epsilon \ln |k\tau_\ast|$, whereas
	the correction quadratic in logarithms was typically proportional to
	$\sim (\epsilon \ln |k\tau_\ast|)^2$, and so on.%
		\footnote{There is no general proof that
		the power series in $\ln |k\tau_\ast|$ always orders itself
		in this form, and according to a general theorem
		due to Weinberg
		(to be discussed in \S\ref{sec:weinberg})
		there may be cases where it does not.
		In many practical examples where the slow-roll
		approximation applies, however, this structure appears.
		In these expressions, the symbol `$\sim$' is used to mean that
		the coefficient of $\ln|k\tau_\ast|$ is a quantity
		of order $\epsilon$ (but not necessarily $\epsilon$ itself),
		wheras the coefficient of $(\ln|k\tau_\ast|)^2$ is of order
		$\epsilon^2$, and so on.}
	One could therefore expect the series to break down
	as a predictive instrument when $|\ln|k\tau_\ast|| \sim
	\epsilon^{-1}$. This occurs when the mode $k$ is of order
	$\epsilon^{-1}$ e-folds outside the horizon.
	
	Taking into account the construction of Gong \& Stewart, and bearing in
	mind that $\tau_\ast$ is the time at which we wish to evaluate
	the correlation functions, it seems clear that there is nothing
	mysterious about the appearance of powers of $\ln |k\tau_\ast|$.
	Such terms merely express that correlations evolve outside the horizon.
	This evolution is mainly driven by the evolving classical background
	field \cite{Zaldarriaga:2003my},
	but we shall see in \S\ref{sec:dynamical} that
	intrinsically quantum contributions can also be present.
	We conclude that, as in other applications of quantum field theory,
	properties of the background play a role in interpreting the
	physics of infrared divergences.
	In the present case,
	correlation functions can be evaluated at any time of interest
	provided the series of logarithms can be summed or ignored.
	Therefore,
	for observational purposes the key question is the
	value of $\epsilon$ when modes of
	interest leave the horizon.
	We would typically wish to evaluate correlation functions at the end of
	inflation, where initial conditions for the subsequent evolution must be
	set.
	If $\epsilon \gtrsim 10^{-2}$, the end of inflation may be uncomfortably
	close to the era when $|\epsilon \ln |k \tau_\ast|| \sim 1$.

	We will return to this question in \S\ref{sec:time},
	where we will argue that a reasonable prescription exists
	for handling time-dependent logarithms which does not
	impair our ability to extract a predictive initial condition
	even if $|\epsilon \ln | k \tau_\ast | | \sim 1$ at the time
	of interest.
	We can already note one obvious strategy.	
	If only logarithmic divergences are present, then
	better control over the power series may be achieved by ``resumming''---%
	that is, accounting for the contribution of---%
	all terms at the same order as the \emph{leading logarithms},
	which are terms of the form $(\epsilon \ln |k\tau_\ast|)^n$ for
	all $n \geq 0$
	\cite{Seery:2007wf}. For this purpose the principal tool is the
	renormalization group equation. Such an analysis has been carried
	out by Burgess, Holman, Leblond \& Shandera
	\cite{Burgess:2009bs},
	whose method will be outlined in \S\ref{sec:time}.
	Unfortunately, this method
	leaves open the question of whether all divergences
	in the $|k \tau_\ast| \rightarrow 0$ limit are comparatively
	tame logarithms,
	or if more aggressive behaviour can occur.
	
	This question was taken up by Weinberg in 2005
	\cite{Weinberg:2005vy}, and later
	by Chaicherdsakul \cite{Chaicherdsakul:2006ui}.
	Weinberg was able to prove that in many inflationary models,
	but not all, the worst divergences would be logarithmic.
	In a later publication, this theorem was extended to fields
	of higher spin
	\cite{Weinberg:2006ac}.
	What of the possibility of faster divergences,
	which are apparently allowed
	by the analysis of Sasaki {\etal} \cite{Sasaki:1992ux}?
	These would grow like powers of the scale factor, $a(t)$,
	corresponding to powers rather than logarithms of $|k \tau_\ast|$.
	In many cases such aggressive growth would induce a failure
	of predictivity long before the end of inflation.
	Also, whereas the coefficient which accompanies a logarithmic
	divergence is physically meaningful, and can be used as an input
	to renormalization group calculations, power-law divergences are
	by contrast mostly meaningless. The late-time physics whose presence
	they signal must be found elsewhere: it cannot usually be
	distilled from the properties of the divergences themselves.
	Although Weinberg's theorem identifies a class of models
	containing interactions which may give rise to such fast divergences,
	it does not appear that any previously proposed
	inflationary model makes essential use of such interactions.
	For this reason they have not yet received much attention.
	
	\subsection{Box-cutoff logarithms}
	\label{sec:zoo-box}
	
	Until non-linear questions became pressing,
	inflationary perturbation theory was dominated by
	the traditional Lifshitz approach
	\cite{Lifshitz:1945du,Brandenberger:1993zc},
	in which one separates each field into a background $\phi$
	and perturbation $\delta\phi$.
	At second order and above such calculations become more arduous,
	although they have now been carried to an impressive degree of
	refinement \cite{Hwang:2005he,Hwang:2007wq,Christopherson:2009fp}.
	
	The separate universe principle is
	an alternative to the Lifshitz approach.
	According to this principle, a volume of spacetime containing
	a background field $\phi$ and perturbation $\delta\phi$ behaves
	(on scales sufficiently large that
	gradients may be neglected)
	just like an unperturbed universe containing the homogeneous field value
	$\phi + \delta\phi$.
	If we solve for an arbitrary quantity $U$
	with initial conditions set by the value
	of a background field at time $\tau_\ast$
	in the unperturbed universe,
	we can determine the values taken by this
	quantity on superhorizon scales in the perturbed
	universe using the trivial identity
	\begin{equation}
		\delta U(\tau) =
			U(\tau, \phi_\ast + \delta \phi_\ast) -
			U(\tau, \phi_\ast) ,
		\label{eq:separate}
	\end{equation}
	where $\tau$ is the time at which we wish to evaluate $\delta U$.
	This might typically be at the end of inflation or later.
	Whatever time we choose,
	Eq.~\eref{eq:separate} allows correlations of $\delta U$ to be
	computed provided we know the correlation functions of
	$\delta \phi$ at time $\tau_\ast$. For example, the first contribution
	to the two-point function of $\delta U$ is
	$\langle \delta U(\tau) \delta U(\tau) \rangle =
	U_{,\phi_\ast}^2(\tau) \langle \delta \phi \delta \phi \rangle_\ast$,
	where $U_{,\phi_\ast}$ denotes the partial derivative of $U$
	with respect to $\phi_\ast$.
	
	Higher-order terms present in 
	Eq.~\eref{eq:separate} imply that other contributions
	must exist. Let us restore spatial arguments for clarity.
	Even if $\delta\phi_\ast$ has Gaussian statistics at
	time $\tau_\ast$, there will be contributions of the form
	\begin{eqnarray}
		\fl\nonumber
		\langle \delta U(\tau, \vect{x}_1) \delta U(\tau, \vect{x}_2)
		\rangle
		\supseteq
		\frac{1}{4} U_{,\phi_\ast \phi_\ast}^2 \langle
			\delta \phi(\vect{x}_1) \delta \phi(\vect{x}_1)
			\delta \phi(\vect{x}_2) \delta \phi(\vect{x}_2)
		\rangle_\ast
		\\ \mbox{}
		+ \frac{1}{6} U_{,\phi_\ast} U_{,\phi_\ast \phi_\ast \phi_\ast}
			\langle
			\delta \phi(\vect{x}_1) \delta \phi(\vect{x}_2)
			\delta \phi(\vect{x}_2) \delta \phi(\vect{x}_2)
			+ (\vect{x}_1 \leftrightarrow \vect{x}_2)
			\rangle_\ast ,
	\end{eqnarray}
	where $(\vect{x}_1 \leftrightarrow \vect{x}_2)$ denotes the
	preceding term with $\vect{x}_1$ and $\vect{x}_2$ exchanged
	and the symbol ``$\supseteq$'' means that the correlation
	function contains this contribution among others.
	Indeed,
	these four-point correlations will be accompanied by six-, eight-
	and higher $2n$-point correlation functions for all $n$.
	If $\delta\phi_\ast$ has non-gaussian statistics then even more
	contributions will be generated.
	Zaballa, Rodr\'{\i}guez \& Lyth introduced a set of diagrammatic
	(``Feynman-like'')
	rules designed to keep track of these terms
	\cite{Zaballa:2006pv}.
	Generalized rules were discussed by
	Byrnes, Koyama, Sasaki \& Wands \cite{Byrnes:2007tm}
	and in Ref.~\cite{Seery:2007wf}, but practical calculations
	are rarely of sufficient complexity to require them.	
	
	These contributions depend on the correlations
	among $\delta\phi_\ast$ when $|\vect{x}_2 - \vect{x}_1|$ becomes
	large, which are identified most simply
	in Fourier space.
	We find that
	$\langle \delta \phi^2(\vect{x}_1)
	\delta \phi^2(\vect{x}_2) \rangle_\ast$ receives contributions
	of the form
	\begin{equation}
		\langle \delta \phi^2 (\vect{x}_1)
		\delta \phi^2 (\vect{x}_2) \rangle_\ast
		\supseteq
		\int \frac{\d^3 k}{(2\pi)^3}
		\e{\im \vect{k} \cdot ( \vect{x}_1 - \vect{x}_2 )}
		\int \frac{\d^3 q}{(2\pi)^3}
		P(|\vect{k}-\vect{q}|) P(q) ,
		\label{eq:nonlin}
	\end{equation}
	in which $P(q)$ is the power spectrum of $\delta\phi_\ast$, defined
	by
	\begin{equation}
		\langle \delta \phi(\vect{k}_1) \delta \phi(\vect{k}_2) \rangle_\ast
		= (2\pi)^3 \delta(\vect{k}_1 + \vect{k}_2) P(k_1) .
		\label{eq:ps}
	\end{equation}
	In the limit $|\vect{x}_1 - \vect{x}_2| \rightarrow \infty$, significant
	contributions to the $\vect{k}$-integral in
	Eq.~\eref{eq:nonlin} arise only from the region
	$k \ll |\vect{x}_1 - \vect{x}_2|^{-1}$.
	The behaviour of the integrand in the small-$k$ limit
	depends crucially on
	$P(q)$. For a scale-invariant power spectrum generated during inflation
	$P(q) \sim H^2 / 2q^3$, which gives
	logarithmic singularities in the $\vect{q}$-integral.
	Boubekeur \& Lyth \cite{Boubekeur:2005fj}, and later
	Lyth \& Rodr\'{\i}guez \cite{Lyth:2005fi},
	remarked that because the logarithmic behaviour is softened in the
	limit $q \gg k$, the log-divergent part only receives contributions
	for $q \lesssim k$. Therefore, to a reasonable approximation,
	\begin{equation}
		\langle \delta \phi^2(\vect{x}_1) \delta \phi^2(\vect{x}_2)
		\rangle_\ast
		\simeq
		\left( \frac{H^2}{2\pi^2} \right)^2
		\int \frac{\d^3 k}{2k^3} \;
		(\ln kL)
		\e{\im \vect{k} \cdot (\vect{x}_1 - \vect{x}_2)} ,
		\label{eq:boxlog}
	\end{equation}
	where $L \gg |\vect{x}_1 - \vect{x}_2|$
	is an infrared cutoff. This expression diverges in the
	limit $L \rightarrow \infty$.
	
	It is clear at once that this divergence is not a physical
	prediction. It depends on what is assumed about correlations
	among the perturbations in the infinite volume limit,
	where $|\vect{x}_2 - \vect{x}_1| \rightarrow \infty$.
	Even within the framework we are using, it is easy to see that
	if $P(q)$ has constant red tilt then this logarithmic divergence
	is exchanged for a power law.
	If $P(q)$ has constant blue
	tilt the integral is convergent.
	An exactly logarithmic divergence occurs only for a scale invariant $P(q)$.
	For example,
	if $\delta\phi$ has even a very small mass this will give
	rise to a finite correlation length, beyond which correlations
	decay exponentially.
	The divergence in Eq.~\eref{eq:boxlog} would then be absent.
	We will see later that there are good physical reasons to believe this
	is what should happen in practice, at least in certain theories, and
	some evidence from concrete calculations that it does.
	
	In any case
	we should also recognize that the framework we have been discussing
	is inadequate for the description of correlations on very large scales.
	Why is this?
	To calculate the power spectrum from a model of inflation we must
	typically assume that the background field
	$\phi$ is spatially homogeneous and depends only on time,
	whereas the spatially dependent fluctuation $\delta \phi$
	satisfies $|\delta\phi| \ll |\phi|$ everywhere.
	Although this is reasonable on scales not too much larger than
	the de Sitter horizon, it need not happen that every
	field admits such a decomposition as we pass to the
	infinite volume limit.
	If this is the case we should set the scale $L$ to be conservatively
	smaller than the largest scale for which a spatially homogeneous
	background $\phi$ can be found.
	The calculation makes sense within this ``box.''
	If we are forced
	to discuss correlations on larger scales, they will have to
	be determined by patching together a mosaic of boxes
	in which the homogeneous background field may take different values.
	This point of view was advocated by Boubekeur \& Lyth
	\cite{Boubekeur:2005fj}, and later refined by
	Byrnes, Koyama, Sasaki \& Wands \cite{Byrnes:2007tm},
	Lyth \cite{Lyth:2006gd,Lyth:2007jh} and other authors
	\cite{Seery:2007wf,Bartolo:2007ti,Enqvist:2008kt,Kohri:2009ac}.
	
	This can not be the whole story, because
	Eq.~\eref{eq:boxlog} shows that the mosaicking procedure
	leaves behind terms involving $\ln kL$.
	These logarithms depend on the arbitrary
	scale $L$.
	Assuming the physics of the scalar field zero-mode
	to be quasi-classical on very large scales,
	the missing element is an accurate map of the average scalar field
	value within each box of the mosaic.
	This map naturally depends on the scale $L$,
	but the $L$ dependence of physical quantities cancels
	when we study correlations within the mosaic as a whole.
	These issues will be discussed in more detail in
	\S\ref{sec:space}.
	We see again that the appearance
	of infrared divergences is connected with the existence of non-trivial
	background configurations. In the present case the map could consist
	of an inventory of boxes, pairing each box with the average field
	value within.
	Alternatively, in a theory such as inflation
	where predictions are essentially statistical, it could
	simply consist of a probability distribution for the average field
	value, taken over the ensemble of boxes.
	
	\subsection{Logarithms from new physics}
	\label{sec:zoo-new}
	
	In the previous two sections we have emphasized the role of the background
	field configuration in generating infrared divergences:
	When determining correlations among small fluctuations we take the
	background to be homogeneous and approximately time independent.
	In practice it may be neither, leading to the emergence of
	compensating logarithms.
	
	There is another source of logarithmic corrections
	which is unconnected
	with the background field configuration.
	Suppose we define some quantum field theory at a scale $\mu$.
	In using this field theory to compute correlations in vacuum---%
	where there is no background at all---%
	we are familiar with the appearance of
	``large logarithms'' of the form $\ln E/\mu$,
	where $E$ is an energy scale characteristic of the correlation in
	question.
	Corrections of this sort occur in any quantum field theory, and
	the field theories we use to compute inflationary correlations are
	no exception.
	Such logarithms were studied by Weinberg
	\cite{Weinberg:2005vy,Weinberg:2006ac}, who computed
	the correction induced by loops of $N$ different spectator
	fields in a model where inflation is driven by the
	vacuum energy associated with a single scalar field.
	
	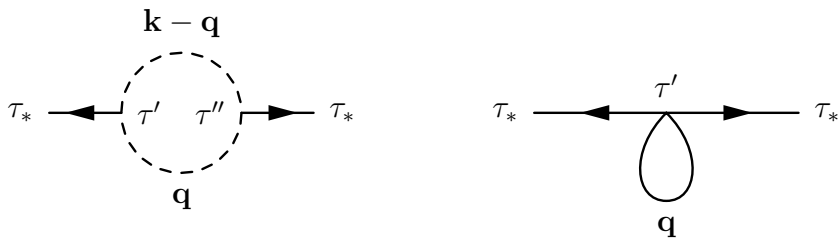
\begin{figure}
		\hfill
		\begin{fmfgraph*}(100,80)
			\fmfpen{thin}
			\fmfleft{l}
			\fmfright{r}
			\fmf{fermion}{v1,l}
			\fmf{dashes,left,tension=0.3,label=$\vect{k}-\vect{q}$}{v1,v2}
			\fmf{dashes,left,tension=0.3,label=$\vect{q}$}{v2,v1}
			\fmf{fermion}{v2,r}
			\fmfv{label=$\tau_\ast$}{l}
			\fmfv{label=$\tau_\ast$}{r}
			\fmfv{label=$\tau'$,label.angle=0}{v1}
			\fmfv{label=$\tau''$,label.angle=180}{v2}
		\end{fmfgraph*}
		\hfill
		\begin{fmfgraph*}(100,80)
			\fmfpen{thin}
			\fmfleft{l}
			\fmfright{r}
			\fmf{fermion}{v,l}
			\fmf{fermion}{v,r}
			\fmf{plain,left,tension=1.0,label=$\vect{q}$}{v,v}
			\fmfv{label=$\tau_\ast$}{l}
			\fmfv{label=$\tau_\ast$}{r}
			\fmfv{label=$\tau'$,label.angle=90}{v}
		\end{fmfgraph*}
		\hfill
		\mbox{}
		\caption{One-loop corrections to the power spectrum
		of an active scalar field. In the left-hand diagram
		a loop of spectator scalar fields, represented by dashed lines,
		corrects the two-point function of an active field,
		represented by solid lines. This loop was computed by
		Weinberg \cite{Weinberg:2005vy,Weinberg:2006ac}. An error in
		the numerical coefficient was corrected by Adshead, Easther
		\& Lim \cite{Adshead:2008gk}.
		In the right-hand diagram, a self-loop corrects the same
		two-point function of an active scalar, first computed
		in Ref.~\cite{Seery:2007we} and again in Ref.~\cite{Adshead:2008gk}.
		In Minkowski space this diagram would factorize, leaving
		a scale-free integral over the loop momentum $\vect{q}$.
		The non-trivial time dependence of de Sitter endows
		the loop with a scale, of the form of Eq.~\eref{eq:uvlog}.
		This diagram is the leading correction when self-loops are
		included. It would be accompanied by self-loops of the same
		form as the left-hand diagram, which are suppressed by powers of
		the slow-roll parameter $\epsilon$. For spectator fields there
		is no contribution from the right-hand diagram, so the left-hand
		loop is the leading term.
		\label{fig:oneloop}}
	\end{figure}

	[See Fig.~\ref{fig:oneloop}.
	In contrast to Minkowski space,
	each external leg is evaluated at the \emph{same} late time
	$\tau_\ast$ and therefore
	these diagrams should be read from the middle to the outside.
	The interior of each diagram functions rather like an instanton.%
		\footnote{The conventional prescription for calculating
		correlation functions in an interacting vacuum can be
		related to the Hartle--Hawking state \cite{Clunan:2009ib}.}
	Breaking the left-hand diagram down the middle,
	\emph{four} quanta of the spectator field nucleate
	below the de Sitter horizon, and propagate until one pair
	annihilates at time $\tau'$ and a second pair at time $\tau''$.
	The product of each annihilation is a quantum of the active
	field, which propagates freely until the surface $\tau = \tau_\ast$
	(represented by the outgoing arrows).
	The shared history of these two particles generates
	a correlation. In an alternative and equally acceptable
	interpretation, a pair of active quanta nucleate. One particle
	propagates directly to the surface $\tau = \tau_\ast$,
	whereas the second spontaneously fluctuates into
	a pair of spectator quanta at time $\tau'$, and at a later
	time $\tau''$
	coalesce to form another quantum of the active field
	before propagating to the final surface.
	The right-hand diagram can be interpreted
	similarly, with the self-loop either providing a correction
	to the nucleation event, or dressing the propagation of a
	single quantum on its route to the surface $\tau = \tau_\ast$.]

	As in Minkowski space, these logarithms arise from ultraviolet
	divergent momentum integrals which we can study
	by cutting off the integral for $q > \Lambda$.
	For a correlation dominated by wavenumbers of order $k$,
	the typical integrals with which we are confronted take a form similar to
	\begin{equation}
		\int_{|q| < \Lambda} \frac{\d^3 q}{(2\pi)^3} \frac{k}{q^3}
		\frac{(\vect{k} \cdot \vect{q})^2}{|\vect{q} + \vect{k}|^2}
		= \frac{k^3}{6\pi^2} \ln \frac{\Lambda}{k} +
		\mathrm{finite} ,
		\label{eq:uvlog}
	\end{equation}
	where $\vect{q}$ is the comoving three-momentum which circulates in
	the loop, and
	``finite'' denotes terms which do not diverge in the limit
	$\Lambda \rightarrow \infty$.%
		\footnote{Other infinite terms may be present, which
		diverge like powers in the limit $\Lambda \rightarrow \infty$,
		rather than logarithms. We are tacitly assuming that these
		divergences can be absorbed by renormalization-group running
		of the masses and couplings.}
	It is the appearance of
	$\vect{k}$ in the combination $|\vect{q} + \vect{k}|$ which gives
	the logarithm a non-trivial $k$ dependence, and
	by the usual arguments we can be assured
	that the coefficient of the logarithm is independent of the
	precise ultraviolet regulator we choose. This does not make
	Eq.~\eref{eq:uvlog} entirely unambiguous because its interpretation
	depends on the meaning we assign to $\Lambda$, as we shall discuss
	shortly. Whatever its meaning, we see the emergence of
	the structure $E/\mu$, with $E = k$ characterizing the scale of
	the correlation, and $\Lambda = \mu$ representing the point at which
	we define the theory.
	
	These logarithms are a different species to the infrared logarithms
	of \S\S\ref{sec:zoo-time}--\ref{sec:zoo-box}, which diverged
	for fixed $k$ when the calculation was taken to occur in a box of
	infinite spatial or temporal extent.%
		\footnote{Following Weinberg's computation of loop corrections of the
		form~\eref{eq:uvlog},
		and others,
		Ref.~\cite{Seery:2007wf} combined
		the effect of secular and box-cutoff logarithms
		with that of an ultraviolet logarithm at the same order
		of perturbation theory. In this work, the scale $\Lambda$
		which accompanies the ultraviolet logarithm was erroneously identified
		with the scale $L$ which accompanies the box-cutoff logarithm.
		This error was corrected from \texttt{v3} onwards of the
		arXiv version of Ref.~\cite{Seery:2007wf}.
		I would like to thank E. Dimastrogiovanni for drawing this
		to my attention.}
	Eq.~\eref{eq:uvlog}
	is finite in this limit, although we will see below that
	quantum loops can also generate infrared divergent terms.
	It is true that Eq.~\eref{eq:uvlog}
	has bad behaviour when $k \rightarrow 0$, but
	a divergence in this limit occurs even at tree-level.%
		\footnote{This divergence arises from our convention
		of normalizing the field modes as Minkowski space oscillators
		deep inside the horizon, but it is not clear any
		physical effect exists because
		the divergence will be
		cut off in any model
		where inflation began at a finite point in the past.
		Alternatively,
		Parker and collaborators have claimed that a different
		procedure is required when extracting observables.
		In a free field theory on Minkowski space, the operator
		product expansion for some local
		functional $\mathcal{O}(x)$
		can be defined via normal ordering,
		\begin{equation}
			\mathcal{O}(x) \mathcal{O}(x')
			=
			F( x - x' ) \; {: \! \mathcal{O}^2(x) \! :} \, \mbox{} + \,
			G( x - x' ) \, + \, \cdots ,
		\end{equation}
		where $F$ and $G$ may be singular as $x' \rightarrow x$,
		and `$\cdots$' denotes terms which are finite in this limit.
		The difference
		between $\mathcal{O}(x) \mathcal{O}(x')$ and
		the composite operator $: \! \mathcal{O}^2(x) \! :$
		(which has a finite vacuum expectation value)
		accounts for the singularities of the Wightman functions
		at coincidence.
		For an interacting field theory on curved
		spacetime the normal-ordered product $: \! \mathcal{O}^2(x) \! :$
		must be replaced by a more general definition of a renormalized
		composite operator, $\mathcal{O}^2_R(x)$.
		Parker {\etal} propose that
		such operators be defined by an adiabatic subtraction
		procedure rather than normal ordering.
		(In principle, the curved space operator product expansion
		of Hollands \& Wald \cite{Hollands:2008ex,Hollands:2008vx,
		Hollands:2001fb,Hollands:2001nf,Hollands:2002ux}
		could be applied instead.)
		In their prescription, contributions to the classical
		power spectrum should be computed
		at horizon exit
		using the Fourier modes of $\mathcal{O}^2_R(x)$ rather than
		$\mathcal{O}(x) \mathcal{O}(x')$,
		as is conventionally done.
		The adiabatic subtraction procedure which makes the renormalized
		composite operator $\mathcal{O}_R^2$ well-defined in the ultraviolet
		also
		entails subtraction of the low-$k$ modes
		which give infrared
		divergences, at least in the massless case
		\cite{Parker:2007ni,Agullo:2010ui}.
		At the time of writing, most authors prefer to define the power
		spectrum using the Fourier modes of
		$\mathcal{O}(x) \mathcal{O}(x')$.}
	We nevertheless include these
	ultraviolet logarithms in this discussion of infrared effects
	because they alert us to the fact that large-scale (that is,
	infrared) structure in the background fields depends on
	physics at large values of the Hubble rate, $H$ (that is,
	in the ultraviolet).
	This feature of gravity, which swaps ultraviolet and infrared
	physics, is known from previous applications of the
	holographic renormalization group to domain wall spacetimes
	which are asymptotically anti de Sitter
	\cite{Balasubramanian:1998sn,
	Bianchi:2001de,Bianchi:2001kw}
	or de Sitter \cite{Larsen:2002et,Larsen:2003pf,vanderSchaar:2003sz}.
	We will return to this question when we revisit the
	mosaicking prescription 
	outlined in \S\ref{sec:zoo-box},
	because it is clear that the large-scale structure of the mosaic
	will depend on what is assumed about the matter field theory
	in the ultraviolet.
	
	We must still decide what meaning should be attached to $\Lambda$.
	In calculating loop corrections,
	Weinberg \cite{Weinberg:2005vy,Weinberg:2006ac}
	made use of dimensional regularization.
	This entails an analytic continuation
	to $3 + \epsilon$ spatial dimensions, after which a limit
	must be extracted by dropping singular terms
	as $\epsilon \rightarrow 0$ \cite{Leibbrandt:1975dj}. Later,
	Chaicherdsakul \cite{Chaicherdsakul:2006ui} and
	Adshead, Easther \& Lim \cite{Adshead:2008gk,Adshead:2009cb}
	employed the same technique.
	Because $\Lambda$ does not appear explicitly in this method,
	being hidden in subtractions associated with the pole at
	$\epsilon = 0$,
	these papers quoted the loop correction as a multiple of $\ln k$
	but left the scale $\Lambda$ implicit.
	Eq.~\eref{eq:uvlog} used an alternative procedure, imposing
	a sharp limit on the comoving momenta which contribute to each
	integral.
	This method
	was employed in Ref.~\cite{Seery:2007we},
	and later by Dimastrogiovanni \& Bartolo
	\cite{Dimastrogiovanni:2008af},
	and gives results in agreement with dimensional regularization.

	In either approach, $\Lambda$ must be interpreted as a comoving scale.
	van der Meulen \& Smit remarked that because comoving scales
	are not physical, logarithms of the form
	$\ln \Lambda/k$ were not easy to interpret
	\cite{vanderMeulen:2007ah}. It had previously
	been observed in Ref.~\cite{Seery:2007we} that if
	one takes $\Lambda$ to be associated with a fixed physical scale
	$\Lambda_P$, then at an arbitrary time
	the corresponding comoving scale is $\Lambda =
	\Lambda_\mathrm{P} a$ where $a$ is the scale factor.
	But at which time should the cutoff be evaluated?
	Senatore \& Zaldarriaga argued that an appropriate choice would
	be the time of horizon crossing
	\cite{Senatore:2009cf}, for which $a = k / H_k$ if
	$H_k$ is the value of the Hubble parameter as the wavenumber
	$k$ crosses the horizon,
	at time $\tau = \tau_k$.
	If so we could conclude that
	Eq.~\eref{eq:uvlog}, and other similar integrals,
	should be replaced by
	\begin{equation}
		\int \frac{\d^3 q}{(2\pi)^3} \frac{k}{q^3}
		\frac{(\vect{k} \cdot \vect{q})^2}{|\vect{q} + \vect{k}|^2}
		= \frac{k^3}{6\pi^2} \ln \frac{\LambdaPhys}{H_k} +
		\mathrm{finite} .
		\label{eq:uvlog-sz}
	\end{equation}
	A similar conclusion had been reached earlier, by a different
	method, in Ref.~\cite{Seery:2008ms}.
	This would
	give a result in precise analogy with
	the logarithm $\ln E/\mu$ encountered in scattering calculations,
	but seems incompatible with the use of dimensional regularization.
	However, Senatore \& Zaldarriaga went on to argue
	that the free-field propagators which appear in the Feynman rules
	should also be calculated by
	analytic continuation of the Mukhanov equation to $3+\epsilon$
	spatial dimensions.
	This was not attempted in Refs.~\cite{Weinberg:2005vy,Weinberg:2006ac,
	Chaicherdsakul:2006ui,Adshead:2008gk}.
	Analytically continuing in this way
	generates extra terms which enable
	dimensional regularization to reproduce~\eref{eq:uvlog-sz}.
	At the present time it seems unclear whether the horizon-crossing
	cutoff can be obtained from a more fundamental principle, or whether
	one must simply adopt it as a prescription giving reasonable results
	consistent with the approximate de Sitter symmetry.%
		\footnote{Senatore \& Zaldarriaga observed that the choice
		$a = k/H_k$ in the comoving cutoff was equivalent to
		including only quanta beneath the cutoff at
		the earliest of the times $\tau'$ and $\tau''$
		in the left-hand diagram of Fig.~\ref{fig:oneloop}.
		In a more complicated diagram, one would choose the earliest
		such time.
		Essentially the same argument was given in Ref.~\cite{Seery:2008ms},
		which discussed
		the influence of quanta which redshift under the horizon
		at times subsequent to $\tau'$.}
	It is also unclear how this form of dimensional regularization
	should be extended to momentum integrals arising from field
	redefinitions, which are to be discussed below. At one loop
	these can be dealt with using a fixed momentum cutoff,
	so this ambiguity is not yet pressing.
	
	\parag{Corrections to the power spectrum.}
	For convenience, some results on loop corrections will be
	collected in this section.
	
	Weinberg gave his calculation in the comoving gauge, where
	the role of active scalar field was taken by the comoving
	curvature perturbation, $\zeta$.
	Adshead, Easther \& Lim worked in the uniform curvature gauge,
	where the action describing a 3-point contact interaction
	between an active field $\phi$ and a collection of $\Niso$
	spectator fields $s^\alpha$ can be written
	\cite{Seery:2005gb}
	\begin{equation}
		\fl
		S_3 \supseteq \int \d^4 x \; \frac{a^2}{2 \Mp^2}
		\frac{\dot{\phi}}{H}
		\left(
			\frac{1}{2a} \frac{\delta L}{\delta \phi}
				\left[
					\partial^{-2} \left( s_\alpha \partial^2 s_\alpha \right)
					- \frac{1}{2} s_\alpha s_\alpha
				\right] 
			- \frac{1}{2} \delta \phi s'_\alpha s'_\alpha
			- \partial^{-2} \delta\phi' s'_\alpha \partial^2 s_\alpha
		\right),
	\end{equation}
	where $\d^4 x = \d^3 x \, \d \tau$,
	summation of repeated $\alpha$ indices is implied,
	a prime $'$ denotes a derivative with respect to conformal time
	and an overdot denotes a derivative with respect to cosmic time.
	In this equation, following the notation of Maldacena
	\cite{Maldacena:2002vr},
	$\delta L / \delta \phi$ represents the first-order
	equation of motion for $\phi$.
	This term can be removed by a field redefinition%
		\footnote{This procedure is notationally misleading, because
		$\delta L / \delta \phi$ is zero by construction
		(at least to leading order)
		when evaluated on a propagator
		and therefore gives no contribution whether we subtract it or not.
		The $\delta L / \delta \phi$
		term is a proxy for boundary terms, which following the
		notation of Maldacena \cite{Maldacena:2002vr} have not been
		written explicitly, which do \emph{not} vanish
		on solutions to the equations of motion
		\cite{Seery:2005gb}.
		It was observed in Ref.~\cite{Seery:2006tq} that
		the boundary terms and $\delta L / \delta \phi$ terms are
		both removed by the field definition, allowing this notational
		trick to make sense.}
	\begin{equation}
		\delta \phi = \sigma - \frac{\dot{\phi}}{4\Mp^2 H}
		\left( \partial^{-2} \left( s_\alpha \partial^2 s_\alpha \right) -
		\frac{1}{2} s_\alpha s_\alpha \right) .
	\end{equation}
	The rules for evaluating correlation functions such as
	$\langle \sigma \sigma \rangle$ are discussed in the review paper
	by Koyama elsewhere in this volume
	\cite{Koyama:2010xj}. One finds the power spectrum
	of $\sigma$ to be
	\begin{equation}
		P^{(\sigma)}_\ast(k) = \frac{H_\ast^2}{2k^3}
		\left(1 - \frac{\Niso \epsilon_\ast}{30\pi^2} \frac{H_\ast^2}{\Mp^2}
		\ln \frac{\LambdaPhys}{H_\ast} + \cdots
		\right)
	\end{equation}
	where now the time of observation, $\tau_\ast$, is chosen to be
	almost immediately after
	the time of horizon crossing, $\tau_\ast \approx \tau_k$.
	
	To obtain the two-point correlation function
	of $\delta \phi$, the effect of the field redefinition
	must be reversed. One finds
	\begin{eqnarray}
		\fl\nonumber
		\langle \delta \phi(\vect{k}_1) \delta \phi(\vect{k}_2) \rangle_\ast
		\supseteq
		\langle \sigma(\vect{k}_1) \sigma(\vect{k}_2) \rangle_\ast
		\\ \nonumber
		\hspace{-1cm}
		\mbox{}
		- \frac{\dot{\phi}_\ast}{4 \Mp^2 H_\ast}
		\Big\langle \sigma(\vect{k}_1)
			\left(
				\partial^{-2} \left(
					s_\alpha \partial^2 s_\alpha
				\right) - \frac{1}{2} s_\alpha s_\alpha
			\right)_{\vect{k}_2}
		\Big\rangle_\ast
		 +
		( \vect{k}_1 \leftrightarrow \vect{k}_2 )
		\\
		\hspace{-1cm}
		\mbox{}
		+ \frac{\dot{\phi}^2_\ast}{16 \Mp^4 H_\ast^2} \Big\langle
			\left(
				\partial^{-2} \left(
					s_\alpha \partial^2 s_\alpha
				\right) - \frac{1}{2} s_\alpha s_\alpha
			\right)_{\vect{k}_1}
			\hspace{-2mm}
			\left(
				\partial^{-2} \left(
					s_\beta \partial^2 s_\beta
				\right) - \frac{1}{2} s_\beta s_\beta
			\right)_{\vect{k}_2}
		\Big\rangle_\ast ,
	\end{eqnarray}
	where $(\vect{k}_1 \leftrightarrow \vect{k}_2)$ denotes the previous
	term with $\vect{k}_1$ and $\vect{k}_2$ exchanged, and
	$(AB)_{\vect{k}}$ is the convolution of $A$ and $B$ with
	argument $\vect{k}$.
	In evaluating this expression we encounter three-point correlations
	of the form $\langle \sigma s_\alpha s_\alpha \rangle$
	and four-point correlations of the form
	$\langle s_\alpha s_\alpha s_\beta s_\beta \rangle$, but the
	four-point terms give no contribution to
	$\ln (\LambdaPhys/H_\ast)$.
	The three-point correlations can be evaluated using the
	general formula for a three-point function of
	scalar fields given in Ref.~\cite{Seery:2005gb}, yielding
	\begin{eqnarray}
		\fl\nonumber
		\langle \sigma(\vect{p}_1) s_\alpha(\vect{p}_2) s_\beta(\vect{p}_3)
		\rangle_\ast =
		(2\pi)^3 \delta(\vect{p}_1 + \vect{p}_2 + \vect{p}_3)
		\frac{1}{4 \prod_i p_i^3} \frac{H_\ast^4}{\Mp}
		\frac{\dot{\phi}_\ast}{4 \Mp H_\ast} \delta_{\alpha \beta} \\
		\mbox{} \times
		\left( - 8 \frac{p_2^2 p_3^2}{p_t} + p_1(p_1^2 - p_2^2 - p_3^2)
		\right) ,
	\end{eqnarray}
	where $p_t = p_1 + p_2 + p_3$.
	After some calculation, one finds that
	the second term in brackets $(\cdots)$ does not contribute,
	and the power spectrum
	of $\phi$ can be written
	\begin{equation}
		P_\ast(k) = \frac{H_\ast^2}{2k^3} \left(
			1 - \frac{3 \Niso \epsilon_\ast}{40 \pi^2} \frac{H_\ast^2}{\Mp^2}
			\ln \frac{\LambdaPhys}{H_\ast} + \cdots \right).
		\label{eq:spectator-loop}
	\end{equation}
	From this calculation we see that field redefinitions reshuffle
	the coefficient of $\ln (\LambdaPhys/H_\ast)$.
	Since gauge transformations are a form of field redefinition, we must
	expect the numerical coefficient $-3/40$ in
	Eq.~\eref{eq:spectator-loop} to shift under a change of gauge.
	In particular, this will happen
	if we change to comoving gauge and
	obtain the correlation function of the curvature perturbation, $\zeta$.
	Therefore Eq.~\eref{eq:spectator-loop} cannot be compared directly to
	Weinberg's result in Ref.~\cite{Weinberg:2005vy},
	but is an equally good indicator of the magnitude of the loop
	correction.%
		\footnote{A calculation of this loop in the uniform curvature
		gauge was given in Ref.~\cite{Adshead:2008gk}, but
		the contribution from the field redefinition was not included.
		In Weinberg's calculation, given in the comoving gauge,
		a similar redefinition
		was dropped, which changes the answer.
		(A numerical error in Weinberg's calculation was corrected
		by Adshead, Easther \& Lim \cite{Adshead:2008gk}, but these
		authors also did not include the
		contribution from the field redefinition.)
		I would like to thank P. Adshead for
		communicating the outcome of his calculations
		concerning this issue.}
	
	Some time after Weinberg's $\zeta$-gauge calculation
	of this spectator loop \cite{Weinberg:2005vy},
	the self-loop correction to the power spectrum
	of an active scalar field was obtained \cite{Seery:2007we}.%
		\footnote{This calculation depends on a set of Feynman rules for
		the active field, which were obtained in the same reference.
		Unfortunately a sign error was present in one of these rules, which
		led to an incorrect numerical coefficient in the final answer.
		This sign error was detected and corrected in
		Ref.~\cite{Adshead:2008gk}, but owing to some typographical errors
		the final numerical coefficient
		determined by these authors was again given incorrectly.
		The correct coefficient was first given in \texttt{v3} of
		Ref.~\cite{Seery:2007we}.}
	This corresponds to the right-hand diagram of Fig.~\ref{fig:oneloop},
	and was found to give a correction to the power
	spectrum of the active field immediately after horizon crossing
	which amounted to
	\begin{equation}
		P_\ast(k) = \frac{H_\ast^2}{2k^3}
		\left( 1 + \frac{1}{3\pi^2} \frac{H_\ast^2}{\Mp^2}
			\ln \frac{\LambdaPhys}{H_\ast} + \cdots \right) .
		\label{eq:self-loop}
	\end{equation}
	When more than once active species is present one should also
	consider loops
	similar to the left-hand diagram of Fig.~\ref{fig:oneloop},
	with active fields circulating in the interior of the diagram.
	In principle these allow the species of active scalars to fluctuate
	into each other, but have not yet been calculated.
	
\begin{figure}
		\hfill
		\begin{fmfgraph*}(100,80)
			\fmfpen{thin}
			\fmfleft{l}
			\fmfright{r}
			\fmf{fermion}{v1,l}
			\fmf{fermion}{v2,r}
			\fmf{boson,left,label=$\vect{k} - \vect{q}$,tension=0.3}{v1,v2}
			\fmf{plain,left,label=$\vect{q}$,tension=0.3}{v2,v1}
			\fmfv{label=$\tau_\ast$}{l}
			\fmfv{label=$\tau_\ast$}{r}
			\fmfv{label=$\tau'$,label.angle=0}{v1}
			\fmfv{label=$\tau''$,label.angle=180}{v2}
		\end{fmfgraph*}
		\hfill
		\begin{fmfgraph*}(100,80)
			\fmfpen{thin}
			\fmfleft{l}
			\fmfright{r}
			\fmf{fermion}{v,l}
			\fmf{fermion}{v,r}
			\fmf{boson,left,label=$\vect{q}$}{v,v}
			\fmfv{label=$\tau_\ast$}{l}
			\fmfv{label=$\tau_\ast$}{r}
			\fmfv{label=$\tau'$,label.angle=90}{v}
		\end{fmfgraph*}
		\hfill
		\mbox{}
		\caption{Graviton loop corrections to the power spectrum of
		a scalar field, calculated by Dimastrogiovanni and
		Bartolo \cite{Dimastrogiovanni:2008af}.
		Unlike the case of scalar loops, the right-hand diagram is
		not slow-roll suppressed compared to the left-hand diagram.
		As before, the interior of these diagrams can be considered
		as a sort of instanton for the nucleation of gravitional
		(wavy lines)
		and scalar quanta (straight lines),
		which propagate to the time of observation
		$\tau_\ast$ on the external legs of the diagram.
		\label{fig:grav}}
	\end{figure}
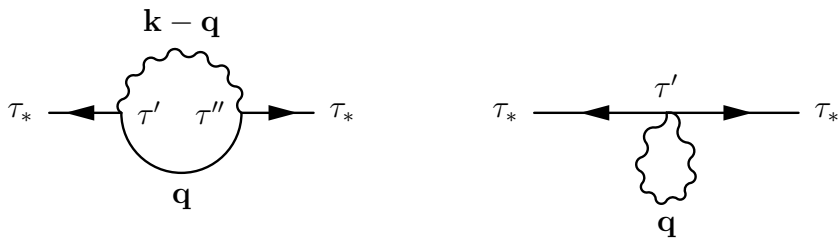

	\parag{Corrections from graviton loops.} Dimastrogiovanni \& Bartolo
	obtained corrections from graviton loops \cite{Dimastrogiovanni:2008af},
	shown in
	Fig.~\ref{fig:grav}.
	Before discussing these,
	consider the self-loop correction for an active scalar field.
	Wick contraction of an external field operator with
	an operator in the interior of the loop
	endows the right-hand diagram of Fig.~\ref{fig:oneloop}
	with a dependence on the external wavenumber.
	This happens because operators like
	$\partial^{-2}$ applied to the interior of the
	loop generate integrals similar to those in Eq.~\eref{eq:uvlog},
	containing $|\vect{k} + \vect{q}|$ in the denominator.
	In the right-hand diagram of Fig.~\ref{fig:grav}
	the graviton pair interior to the loop must contract among themselves,
	and likewise the external fields. Therefore no dependence
	on the external wavenumber is generated, and this diagram factorizes
	into an overall renormalization of the tree-level two-point function,
	\begin{equation}
		P_\ast(k) =
		\frac{H_\ast^2}{2k^3}
		\left( 1 - \frac{1}{2\pi^2}
			\frac{H_\ast^2}{\Mp^2}
			\ln (\LambdaPhys a_\ast L) + \cdots
		\right) ,
		\label{eq:gravbubble}
	\end{equation}
	where $L$ is an infrared cutoff
	of the type which appears in box-cutoff
	logarithms, as discussed in \S\ref{sec:zoo-box}.
	For the left-hand diagram, one finds%
		\footnote{Dimastrogiovanni \& Bartolo quoted their answer in
		a different form, but Eq.~\eref{eq:gravloop} gives the
		only relevant terms in the limit $|k\tau_\ast| \rightarrow 0$.
		Dimastrogiovanni \& Bartolo found a slightly different coefficient
		for each logarithm.
		As this paper was being completed, a recalculation of the
		graviton
		loop appeared in Ref.~\cite{Giddings:2010nc}, which agrees with
		the result given here.
		(P. Adshead \& E. Dimastrogiovanni, personal communications.)}
	\begin{equation}
		P_\ast(k) =
		\frac{H_\ast^2}{2k^3}
		\left( 1 +
			\frac{1}{6 \pi^2}
			\frac{H_\ast^2}{\Mp^2}
			\ln \frac{\LambdaPhys}{H_\ast} +
			\frac{1}{2\pi^2} \frac{H_\ast^2}{\Mp^2} \ln k L + \cdots
		\right) ,
		\label{eq:gravloop}
	\end{equation}
	where ``$\cdots$'' denotes terms which vanish at late times
	together with any subleading contributions.
	These examples are interesting because they
	show that
	quantum loops can generate ``box-cutoff'' effects in addition to
	``new physics'' logarithms,
	which was suggested but not demonstrated in Ref.~\cite{Lyth:2006qz}.
	It is not yet clear how to interpret the association of
	a $\ln kL$ term with a loop containing spin-2 gravitational quanta,
	which lead to traceless deformations of the metric.
	However, these terms all cancel in the total one-loop correction,
	as we now explain.
	
	Together, Eqs.~\eref{eq:spectator-loop}, \eref{eq:self-loop}
	and~\eref{eq:gravbubble}--\eref{eq:gravloop}
	give the largest part of the loop correction for
	each active scalar field
	in a theory containing $\Niso$ spectator fields.%
		\setcounter{footnote}{0}%
		\footnote{Adshead, Easther \& Lim
		argued that for active fields only
		\emph{self}-loops of the
		leading-order interaction give physical logarithmic corrections
		\cite{Adshead:2008gk}, for the reasons discussed above
		Eq.~\eref{eq:gravloop}.
		Loop corrections to the power spectrum of one active species
		from other active species are suppressed by slow-roll factors.}
	The terms which
	are neglected by this combination are higher-order in $\epsilon_\ast$
	but do not include compensating factors of $\Niso$.
	First, the $k$-dependent logarithms in
	Eqs.~\eref{eq:self-loop} and~\eref{eq:gravloop} cancel between themselves,
	to leave a logarithm depending only on the ultraviolet and infrared
	regulators.
	Second, this logarithm cancels with that in Eq.~\eref{eq:gravbubble};
	only the contribution of Eq.~\eref{eq:spectator-loop} is left.
	Therefore, to this accuracy, the total loop correction
	is entirely free of the infrared regulator, $L$,
	\begin{equation}
		P_\ast^{\mathrm{total}}(k) = \frac{H_\ast^2}{2k^3} \left(
			1 - \frac{3 \Niso \epsilon_\ast}{40 \pi^2} \frac{H_\ast^2}{\Mp^2}
			\ln \frac{\LambdaPhys}{H_\ast} + \cdots \right) .
		\label{eq:total-loop}
	\end{equation}
	Remarkably,
	this receives contributions from
	the isocurvature fields alone.
	It is not clear whether this cancellation is accidental, or enforced
	by some deeper principle.
	For example,
	in the case of a
	single minimally coupled scalar field in de Sitter space
	(obtained by passage to the limit $\epsilon \rightarrow 0$),
	this absence of one-loop corrections becomes exact and is
	consistent with de Sitter invariance.
	If the cancellation is accidental there seems no reason
	why $L$-dependent
	logarithms should not appear at subleading order in
	$\epsilon$, or at two loops and higher.
	It would be of considerable
	interest to determine whether this is the case.
	
	Burgess {\etal} \cite{Burgess:2009bs}
	emphasized that, if infrared-sensitive contributions occur,
	one should obtain a time independent answer when
	working with
	\emph{physical}
	ultraviolet and infrared cutoffs, $\LambdaPhys$ and $L_{\mathrm{P}}$.
	In their analysis, this was achieved because the ultraviolet and
	infrared terms combined to give
	$\ln \LambdaPhys L_{\mathrm{P}}$.
	If one chooses the infrared cutoff to be comoving---as we are
	doing in this discussion---then de Sitter
	invariance is broken
	and one cannot avoid a time dependent logarithm
	of the form rejected by Senatore \& Zaldarriaga as incompatible
	with eternal inflation. In the present framework, whether a
	similar problem
	exists depends on the
	time-dependent distribution of background field values which
	removes $L$ from Eq.~\eref{eq:total-loop} and similar
	equations. 
	
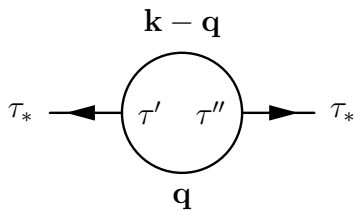
\begin{figure}
		\hfill
		\begin{fmfgraph*}(100,80)
			\fmfpen{thin}
			\fmfleft{l}
			\fmfright{r}
			\fmf{fermion}{v1,l}
			\fmf{fermion}{v2,r}
			\fmf{plain,left,tension=0.3,label=$\vect{k}-\vect{q}$}{v1,v2}
			\fmf{plain,left,tension=0.3,label=$\vect{q}$}{v2,v1}
			\fmfv{label=$\tau_\ast$}{l}
			\fmfv{label=$\tau_\ast$}{r}
			\fmfv{label=$\tau'$,label.angle=0}{v1}
			\fmfv{label=$\tau''$,label.angle=180}{v2}
		\end{fmfgraph*}
		\hfill
		\mbox{}
		\caption{Loop in pure $\phi^3$ theory. A factor of the coupling
		$g$ is present at each vertex.
		\label{fig:cube}}
	\end{figure}

	\parag{$\phi^3$ loop.}
	A somewhat less physical example is the loop in pure $\phi^3$
	theory, which will play an interesting role in the discussion of
	time dependence in \S\ref{sec:time}.
	Taking the interaction to be
	\begin{equation}
		S_3 \supseteq \int \d^4 x \; a^4 \,\frac{g}{3} \delta \phi^3 ,
	\end{equation}
	one arrives at the loop of Fig.~\ref{fig:cube}.
	van der Meulen and Smit calculated this loop, accounting
	for both ultraviolet and infrared effects
	\cite{vanderMeulen:2007ah}. It was later recalculated
	by various groups, since it provides a useful test case
	\cite{Bartolo:2007ti,Seery:2008ms,Burgess:2009bs}. One finds
	\begin{equation}
		\fl
		P_\ast(k) = \frac{H_\ast^2}{2k^3}
		\left( 1 + \frac{g^2}{18 \pi^2} \frac{1}{H_\ast^2}
		(\ln |k \tau_\ast|)^2 \ln k L -
		\frac{4 g^2}{27\pi^2} \frac{1}{H_\ast^2}
		\ln \frac{\LambdaPhys}{H_\ast} + \cdots \right) .
		\label{eq:cubeloop}
	\end{equation}
	
	\parag{$\phi^4$ loop.}
	A loop in $\phi^4$ theory was calculated by
	Tsamis \& Woodard \cite{Tsamis:1997za,Tsamis:2005hd}
	and Petri \cite{Petri:2008ig}, and later by
	Burgess, Holman,
	Leblond \& Shandera \cite{Burgess:2009bs}.
	The interaction is
	\begin{equation}
		S_4 \supseteq \int \d^4 x \; a^4 \; \frac{\lambda}{4!} \delta\phi^4 ,
		\label{eq:phifour}
	\end{equation}
	and the diagram is the same as the right-hand
	loop of Fig.~\ref{fig:oneloop}, although with a different vertex.
	The result is
	\begin{equation}
		P_\ast(k) = \frac{H_\ast^2}{2k^3}
		\left( 1 + \frac{\lambda}{12 \pi^2} \ln k L \ln |k \tau_\ast| +
		\cdots \right) .
		\label{eq:quarticloop}
	\end{equation}
	Other examples of loops have been considered in the literature.
	In Ref.~\cite{Seery:2008ms} a set of loop corrections to the
	two- and three-point functions of an active scalar field
	were calculated in a theory where the scalars coupled to
	the field strength of an Abelian gauge field.
	Although their structure is rather complicated, these loops
	gave results which are not qualitatively different
	to those we have already studied.
	
	\section{Time evolution}
	\label{sec:time}
	
	As in any quantum field theory, the diagrams in
	Figs.~\ref{fig:oneloop}--\ref{fig:cube} are calculated
	by identifying quasi-free propagating modes which 
	travel along the arcs of the diagram and interact at vertices.
	In each calculation discussed above, the quasi-free modes
	behave like massless scalar fields in eternal de Sitter space
	which have two-point correlations of the form
	\begin{eqnarray}
		\fl\nonumber
		\langle
			\delta\phi(\vect{k}_1, \tau_1)
			\delta\phi(\vect{k}_2, \tau_2)
		\rangle
		= (2\pi)^3 \delta(\vect{k}_1 + \vect{k}_2)
		\frac{H_1 H_2}{2k_1^3}
		\\ \mbox{} \times
		\Bigg\{
			\begin{array}{ll}
				(1 - \im k_1 \tau_1)(1 + \im k_2 \tau_2)
				\e{\im (k_1 \tau_1 - k_2 \tau_2)}
				& \mbox{if $\tau_1 < \tau_2$} \\
				(1 + \im k_1 \tau_1)(1 - \im k_2 \tau_2)
				\e{\im (k_2 \tau_2 - k_1 \tau_1)}
				& \mbox{otherwise}
			\end{array}
		,
		\label{eq:massless}
	\end{eqnarray}
	where $H_i = H(\tau_i)$ and $k_i = |\vect{k}_i|$.
	The magnitudes are equal on-shell, where conservation of momentum
	is satisfied, so $k_1 = k_2$.
	The distinction has been maintained in Eq.~\eref{eq:massless}
	because in practical calculations one must often
	calculate off-shell where $k_1 \neq k_2$,
	only taking the on-shell limit at the end of the calculation
	\cite{Seery:2008ms,Seery:2008ax,Adshead:2009cb,
	Chen:2010xk};
	the same is true for higher $n$-point functions where one must
	keep the $k_i$ all distinct.
	Eq.~\eref{eq:massless} is built out of elementary wavefunctions
	proportional to $(1-\im k \tau)\e{\im k \tau}$,
	which asymptotes to unity in the infinite future of
	de Sitter space where $\tau \rightarrow 0^-$.
	Therefore,
	once correlations have been generated using
	Eq.~\eref{eq:massless} they persist into the indefinite future.
	Had we instead taken the quasi-free modes to have a mass of order
	$m$, we would have obtained elementary wavefunctions
	of the form $(-k \tau)^{1/2} a^{-1}(\tau) H_{\nu}^{(2)}(- k \tau)$,
	where $H_{\nu}^{(1,2)}$ are Hankel functions of the first and
	second kind, respectively, and the order of these functions, $\nu$,
	is related to the mass $m$ by
	\begin{equation}
		\nu^2 = \frac{9}{4} - \frac{m^2}{H^2} .
		\label{eq:nu}
	\end{equation}
	For $m \gg H$ this gives the Hankel functions an imaginary order.
	In the limit $\tau \rightarrow 0$ the massive de Sitter wavefunction
	would have an asymptotic expansion proportional to
	\begin{equation}
		\sim (-k \tau)^{3/2 - \nu} \left(
			- \frac{\im 4^\nu \Gamma(\nu)}{\pi} +
			\frac{(-k\tau)^{2 \nu}(1 + \im \cot \pi \nu)}{\Gamma(1+\nu)}
		\right) + \cdots ,
		\label{eq:massive}
	\end{equation}
	where ``$\cdots$'' denotes terms which converge to zero
	more quickly than those which have been written.
	Clearly, correlations are suppressed at late times whenever
	$\nu < 3/2$ and vanish entirely as $\tau \rightarrow 0^-$.
	Having obtained expressions of this form,
	Riotto \& Sloth \cite{Riotto:2008mv} observed that
	Eq.~\eref{eq:massive} expresses nothing more profound than that
	a field of mass $m$ has a proper correlation length of order $1/m$,
	and in the infinite future of de Sitter space
	the cosmological expansion has carried any two spatially distinct
	points far beyond a proper separation of order $1/m$.
	For this reason, correlations
	between spatially separated points must decay.
	
	Weinberg remarked that this mismatch between
	Eq.~\eref{eq:massless} and the decay of correlations in
	the real universe compels us to account for
	mass terms non-perturbatively \cite{Weinberg:2006ac}.
	In principle we could deal with a term in the Lagrangian of the 
	form $m^2 \delta \phi^2$ by including its contribution in the elementary
	wavefunctions of the quasi-free modes, or by counting it among the
	interactions.
	If we include it among the interactions, then we compute
	correlations using Eq.~\eref{eq:massless} and
	are obliged to make the
	implicit assumption that they do not subsequently decay.
	This may yield reasonable results, supposing
	the correlations evolve only slowly,
	provided we do not ask for their
	properties
	too long after horizon crossing.
	Nevertheless, where the correlations truly decay
	this approximation must lead to trouble at late times.
	Since we expect most fields to acquire a non-trivial potential
	except where protected by an exact symmetry,
	we also expect the fluctuations in these fields to develop small
	masses.
	As was explained in \S\ref{sec:zoo-time}, it is our failure to account
	for this time evolution in the background
	fields, and the subsequent decay of correlations,
	which is responsible for the appearance of the secular
	logarithms $\ln |k\tau_\ast|$.
	
	What if a field has a potential containing no mass term, but
	other higher-order interactions?
	This situation was analysed by Senatore \& Zaldarriaga
	\cite{Senatore:2009cf}, who argued that on any potential of this kind
	the background field would still roll down and evolve.
	For this reason, we would expect a mass term to be
	generated for the fluctuations after resumming a sufficient number
	of insertions in the propagator. This would subsequently cause
	correlations
	to decay. Therefore this situation is qualitatively the same
	as the case of a mass term in the potential.
	We will see an explicit example in \S\ref{sec:dynamical}.
	
	\subsection{Weinberg's theorem and related results}
	\label{sec:weinberg}
	
	How fast can we expect correlations to evolve?
	In \S\ref{sec:zoo-time} we discussed a theorem due to Weinberg
	\cite{Weinberg:2005vy}
	which guarantees that only logarithms of $|k \tau_\ast|$
	appear in inflationary correlation functions, and not
	power laws of the form $|k \tau_\ast|^{-n}$ for $n \geq 1$
	which in principle are allowed \cite{Sasaki:1992ux}.
	(Contributions proportional to positive powers of $|k \tau_\ast|$
	are generically present, but contribute nothing at late times.)
	The precise statement of this theorem
	applies to scalar and tensor excitations. In three dimensions,
	we have:
	
	\vspace{3mm}
	\noindent
	\textsf{\textbf{Theorem.}}
	In three space dimensions, power law divergences
	are absent at late times in
	correlations among scalar and tensor quanta,
	provided all interactions in the Lagrangian,
	when written in cosmic time, fall into one of two
	classes:
	\begin{enumerate}
		\item Interactions containing strictly less than one factor
		of $a(t)$.
		\item Interactions which may grow as fast as $a(t)$ (but no
		faster), and which contain only fields rather than
		time derivatives of fields.
	\end{enumerate}

	Weinberg proved this theorem by studying
	commutators among the
	fields and their derivatives at late times, and then counting
	how these commutators could appear in the possible elementary
	interactions.
	This method of proof does not make use of the
	explicit slow-roll, massless solution~\eref{eq:massless}.
	However,
	Senatore \& Zaldarriaga later remarked that a refinement of this
	theorem might be possible \cite{Senatore:2009cf}, because
	the asymptotic estimates obtained by Weinberg \emph{did} assume that the
	fields were massless. In addition, since each elementary interaction
	was considered in isolation, the possibility of cancellations among
	groups of interactions was not considered.
	At the time of writing, these possible refinements have not been
	studied.
	
	In a later publication, Weinberg extended this theorem to Abelian
	and non-Abelian vector fields, and Dirac fermions
	\cite{Weinberg:2006ac}
	(see also Ref.~\cite{Chaicherdsakul:2006ui}).
	These fields have rather better behaviour
	than the scalar fields considered in the original theorem,
	and correlation functions among them typically converge at late times.
	
	\subsection{Resummation by the $\delta N$ formula}
	\label{sec:resum-deltan}
	
	For the remainder of this discussion, we will restrict attention
	to theories satisfying the conditions of Weinberg's theorem.
	Therefore only secular logarithms will appear as
	$\tau_\ast \rightarrow 0$, and not the faster power-law divergences.
	
	Up to this point we have not made our choice of gauge
	explicit, but we have been discussing fluctuations in
	scalar fields.
	This will usually correspond to the uniform curvature
	gauge.
	The quantity whose correlations are accessible to observation in the
	CMB is often taken to be the comoving curvature perturbation, $\zeta$.
	(Indeed, Weinberg's theorem and the spectator loop discussed in
	\S\ref{sec:zoo-new} were given directly in this gauge.)
	The statistical properties of $\zeta$ are determined from a superposition
	of the active fields in the theory.
	Many prescriptions have been employed to obtain this superposition
	\cite{Gordon:2000hv,Malik:2006ir}, but a simple
	method is to observe that $\zeta$ is a local fluctuation in the
	aggregate cosmological expansion, $\zeta = \delta \ln a$, where
	$a$ is the scale factor.
	We can therefore use the separate universe formula,
	Eq.~\eref{eq:separate}.
	One usually defines $\ln a/a_0 = N$, where $N$ is the number of e-folds
	from a reference time where $a = a_0$. We take the initial slice to
	be a uniform curvature hypersurface, and the final slice to be
	a surface of uniform energy density,
	after which we arrive at the $\delta N$ formula
	\cite{Starobinsky:1986fxa,Sasaki:1995aw,Lyth:2005fi},
	reviewed elsewhere in this volume
	by Tanaka, Suyama \& Yokoyama
	\cite{Tanaka:2010km}
	and Wands \cite{Wands:2010af}
	\begin{equation}
		\zeta = \frac{\partial N}{\partial \phi_\ast} \delta \phi_\ast +
			\frac{1}{2} \frac{\partial^2 N}{\partial \phi_\ast^2}
			\delta\phi_\ast^2 +
			\cdots .
		\label{eq:deltan}
	\end{equation}
	Among other things,
	this is a gauge transformation.
	The $\delta N$ formula reproduces those
	terms from the non-linear transformation
	between the uniform curvature gauge and comoving gauge
	\cite{Maldacena:2002vr}
	which do not vanish on superhorizon scales.
	
	Eq.~\eref{eq:deltan} enables the correlation functions of $\zeta$
	to be determined from those of $\delta\phi_\ast$,
	which in turn are computed using the massless propagator~\eref{eq:massless}.
	In the foregoing discussion we argued that
	Eq.~\eref{eq:massless} prevents correlations from decaying
	after they have been generated, and that this corresponds
	to neglecting evolution of the background fields.
	The $\delta N$ formula ameliorates this difficulty.
	Suppose we wish to compute correlations on some scale $k$.
	If we take the initial time, $\tau_\ast$,
	to be shortly after the time of horizon exit associated with $k$,
	then we do not commit a gross error in failing to account for their
	subsequent decay.
	Moreover, the time dependence summarized by $N$ accounts for
	evolution of the background fields as they roll down their potentials.
	Therefore we expect Eq.~\eref{eq:deltan} to incorporate
	the infrared physics whose neglect gave rise to secular logarithms.
	This argument was given in Ref.~\cite{Seery:2007wf}
	and is supported by the analysis of
	van der Meulen \& Smit \cite{vanderMeulen:2007ah},
	although a formal proof is still lacking.
	
	The same conclusion can be reached by considering the logarithms
	themselves. If $\tau_\ast$ is chosen not long after horizon crossing,
	then the terms $\ln |k\tau_\ast|$ are of order unity.
	If the leading term containing $n$ powers of $\ln |k\tau_\ast|$
	occurs in the combination $\sim (\epsilon \ln |k\tau_\ast|)^n$,
	then, not long after horizon crossing,
	all terms containing logarithms will be small compared
	to the tree level whenever $\epsilon \ll 1$.
	Therefore we can neglect all logarithms in using
	Eq.~\eref{eq:deltan} to compute correlations of $\zeta$.
	The effect has been to ``resum'' their effect into the background
	evolution of the scalar fields, which is described by
	the accumulating number of e-folds, $N$.
	For example, in a single-field model of inflation this argument
	reproduces the expected conclusion that $\zeta$ does not evolve.
	But the virtue of the $\delta N$
	method is that it is not restricted to the single field case:
	In a multiple field scenario where $\zeta$ has non-trivial
	evolution, the $\delta N$ formula will correctly describe the
	growth and decay of fluctuations.
	
	There may be problems when using Eq.~\eref{eq:deltan} to obtain
	the correlation functions of $\zeta$ if one wishes to include
	the ``quantum'' logarithms of \S\ref{sec:zoo-new}.
	This procedure was applied in
	Refs.~\cite{Seery:2007we,Seery:2007wf,
	Dimastrogiovanni:2008af,Adshead:2008gk,Seery:2008ms}.
	However,
	as we have already observed,
	the $\delta N$ formula is a gauge transformation, and 
	it was argued in \S\ref{sec:zoo-new} that such transformations
	will typically modify the coefficient of $\ln \LambdaPhys/\Mp$.
	Although this possibility has yet to be investigated in detail,
	the large number of ultraviolet terms discarded by the
	$\delta N$ formula (which should be compared, for example,
	with Eq.~(A.8) of Ref.~\cite{Maldacena:2002vr})
	would apparently make a mismatch likely.
	The $\delta N$ formula retains all relevant infrared
	physics, so we would not expect similar difficulties with
	secular and box-cutoff logarithms.
	
	\subsection{Resummation by the dynamical renormalization group}
	\label{sec:dynamical}
	
	For the purpose of comparison with observation,
	the $\delta N$ method is presumably sufficient
	and allows us to extract initial conditions from typical
	inflationary models without any loss of predictivity.
	In many models of inflation we wish to obtain the
	properties of correlations only 50 to 60 e-folds after they are
	synthesized. In a conventional
	model the only source of appreciable time evolution during
	these 50 to 60 e-folds comes from the classical variation of the
	background.
	Nevertheless, as a point of principle,
	one might prefer a more satisfactory
	resolution.
	Some time after the resummation interpretation of
	$\delta N$ had been proposed,
	Burgess, Holman, Leblond \& Shandera
	remarked that it did not account for the possibility of
	secular time evolution from quantum effects \cite{Burgess:2009bs},
	such as those of Eqs.~\eref{eq:gravloop}, \eref{eq:cubeloop}
	and~\eref{eq:quarticloop}. These terms typically appear
	in combination with the small parameter $(H/\Mp)^2$, and are therefore
	small unless many e-folds have elapsed since horizon crossing.
	
	Burgess {\etal} suggested that these logarithms could be
	resummed using the method of the renormalization group,
	and applied this technique to the quartic loop
	of Eq.~\eref{eq:quarticloop}.
	We briefly review their method.
	The starting point is to introduce an
	arbitrary constant, $c$, which is unity in the
	tree-level power spectrum,
	\begin{equation}
		P_\ast(c,k) = \frac{c H_\ast^2}{2k^3} .
	\end{equation}
	Our strategy is to resum the large logarithms into $c$.
	To do so, rewrite Eq.~\eref{eq:quarticloop} in terms of an arbitrary
	intermediate time scale $\vartheta$,
	\begin{equation}
		\fl
		P_\ast(k) = \frac{c H_\ast^2}{2k^3}
			\left( 1 + \frac{\lambda}{12\pi^2} \ln kL \ln \vartheta
			\right)
			\left( 1 + \frac{\lambda}{12\pi^2} \ln kL \Big[
				\ln |k\tau_\ast| - \ln \vartheta
			\Big] \right) .
	\end{equation}
	Clearly, the combination of overall prefactor and first bracket
	is of the same form as $P_\ast(c,k)$, with $c = c(\vartheta)$.
	We can therefore replace one by the other.
	Whatever the result, it must be independent of the arbitrary time
	$\vartheta$. Applying the familiar method of Gell-Mann and Low
	\cite{GellMann:1954fq}, we find
	\begin{equation}
		\frac{\d c}{\d \vartheta} = \frac{\lambda}{12\pi^2}
		\frac{c}{\vartheta} \ln kL .
		\label{eq:rge}
	\end{equation}
	Solving this differential equation and setting
	$\vartheta = |k \tau_\ast|$, we conclude
	\begin{equation}
		P_\ast(k) =
			\frac{H_\ast^2}{2k^3}
			| k \tau_\ast |^\delta \left(
				1 + \Or[ \epsilon^2 \ln |k\tau_\ast| ]
			\right) ,
		\label{eq:burgess-resum}
	\end{equation}
	where $\delta$ satisfies
	\begin{equation}
		\delta = \frac{\lambda}{12\pi^2} \ln kL .
		\label{eq:delta}
	\end{equation}
	The error in this procedure is of order
	$\epsilon^2 \ln |k\tau_\ast|$. Therefore
	Eqs.~\eref{eq:burgess-resum}--\eref{eq:delta}
	are trustworthy even when $|\epsilon \ln |k\tau_\ast|| \sim 1$,
	giving rise to a resummation of the leading logarithms.
	Subleading logarithms could be included by accounting for
	terms of order $\lambda^2$ or higher in Eq.~\eref{eq:rge}.
	
	Eq.~\eref{eq:burgess-resum} describes decaying correlations.
	On the basis of the foregoing discussion, we expect this behaviour
	to represent the correct infrared physics in a large class of
	theories.
	Indeed,
	comparison with Eq.~\eref{eq:massive} shows that the effect
	of $c(|k\tau_\ast|)$ is to introduce a dependence on
	$|k\tau_\ast|$ comparable to the case of a massive scalar
	field. Making use of Eq.~\eref{eq:nu} and
	recalling that the square of Eq.~\eref{eq:massive}
	yields the power spectrum, we see that
	Eq.~\eref{eq:burgess-resum} is equivalent to a dynamically
	generated mass $\meff$
	\cite{Burgess:2009bs},
	\begin{equation}
		\meff^2 = \frac{\lambda}{8 \pi^2} H_\ast^2 \ln kL .
		\label{eq:npmass}
	\end{equation}
	If $1/k$ and $L$ do not generate an exponential hierarchy and
	$\lambda \sim \Or(1)$, then $\meff$ is of order the Hubble rate.
	We can regard this as a concrete example of the
	argument of Senatore \& Zaldarriaga
	\cite{Senatore:2009cf}
	discussed just before \S\ref{sec:weinberg}.
	
	This conclusion does not apply to every theory.
	Burgess, Holman, Leblond \& Shandera
	observed that the same argument applied to pure $\phi^3$
	theory does not yield a dynamically generated mass.
	The obstruction comes from the power of $\ln |k\tau_\ast|$ in the
	leading term of Eq.~\eref{eq:cubeloop}.
	Ignoring the ultraviolet logarithm proportional to
	$\ln \LambdaPhys/\Mp$, which plays no role here,
	resumming the secular logarithms would give
	\begin{equation}
		P_\ast(k) = \frac{H_\ast^2}{2k^3}
		\exp \left( \frac{g^2}{18\pi^2} \frac{\ln kL}{H_\ast^2}
		(\ln |k\tau_\ast|)^2 \right)
		\left( 1 + \Or [ \epsilon^2 \ln |k\tau_\ast| ] \right) .
		\label{eq:cubedrg}
	\end{equation}
	Since this does not suppress the power spectrum by a positive
	power of $|k\tau_\ast|$ at late times we cannot interpret it
	in terms of an effective value of $\nu$ in
	Eq.~\eref{eq:massive}, and thus an effective mass $\meff$.
	Indeed,~\eref{eq:cubedrg} has rather poor behaviour in the limit
	$|k\tau_\ast| \rightarrow 0$,
	because the exponential diverges there.
	One should not take this behaviour literally.
	Burgess {\etal} included logarithms of cubic order
	which have been neglected here, and which would
	become negative in the limit $\tau_\ast \rightarrow 0^-$.
	Such terms would
	compete with the positive contribtion from
 	$(\ln |k\tau_\ast|)^2$.
	Thus, in pure $\phi^3$
	theory the behaviour of the exponential in the far future
	becomes a delicate question.
	
	This is not difficult to interpret.
	In pure $\phi^3$ theory the Hamiltonian is unbounded below, and
	there is nothing to prevent fluctuations (and their
	correlations) growing indefinitely.
	Burgess {\etal} remarked that this pathological behaviour made
	studying the late-time behaviour of this theory problematic.
	Following the argument of Senatore \& Zaldarriaga
	\cite{Senatore:2009cf},
	we may expect that in realistic theories
	behaviour comparable to~\eref{eq:burgess-resum} is more generic.

	\section{Spatial evolution}
	\label{sec:space}

	Less is known about the resolution of
	what we have called box-cutoff logarithms,
	of the form $\ln kL$, which were left behind in
	\S\ref{sec:zoo-box} after tiling the spatial region of interest
	into a mosaic of boxes.

	\subsection{Mosaicking prescriptions}
	\label{sec:mosaic}

	Our discussion in \S\ref{sec:zoo-box} was based on an
	assumption: that the role of physics at
	scales larger than $L$ was to determine a quasi-classical
	background field configuration
	in each box. A proof of the validity of this assumption
	is not known. What would it entail?
	Consider an expectation value of local operators
	$O_1$, \ldots, $O_n$ evaluated
	at spatial positions $\vect{x}_1$, \ldots, $\vect{x}_n$
	with roughly common separations $| \vect{x}_i - \vect{x}_j| \sim 1/k$
	which are small on the characteristic scale of the mosaic,
	so that $kL \gg 1$.
	We wish to evaluate this correlation function
	at time $\tau_\ast$, just after horizon exit of the wavenumber $k$,
	which can be achieved by averaging against some wavefunctional
	$\Psi[ O ]$ weighting  spatial
	configurations of the $O_i$ \cite{Sasaki:1992ux,Seery:2006wk},
	\begin{equation}
		\langle O_1 (\vect{x}_1) \cdots O_n(\vect{x}_n) \rangle_\ast
		= \int [ \d O ] \; O_1(\vect{x}_1) \cdots O_n(\vect{x}_n)
			\; | \Psi_\ast[ O ] |^2 ,
		\label{eq:expectation-value}
	\end{equation}
	where a subscript `$\ast$' denotes evaluation at $\tau_\ast$.
	Eq.~\eref{eq:expectation-value}
	leads to the in--in or Schwinger path integral
	once each wavefunctional has been
	written as an integral over all histories
	$O(\tau, \vect{x})$ with appropriate boundary conditions at $\tau_\ast$.
	Cut out a box of comoving size~$\sim L$, approximately centred on the
	$\vect{x}_i$. The path integral divides
	into separate integrals over the spatial field configuration
	interior to the box, written $O^-$,
	and the exterior configuration, $O^+$.
	These configurations merge continuously on surfaces of the box,
	denoted $\partial L$.	
	In these terms, Eq.~\eref{eq:expectation-value} can be rewritten
	\begin{equation}
		\fl
		\langle O_1(\vect{x}_1) \cdots O_n(\vect{x}_n) \rangle_\ast
		=
		\int [ \d O^- \, \d O^-_{\partial L} ] \;
		O^-_1(\vect{x}_1) \cdots O^-_n(\vect{x}_n)
		\;
		| \Psi^-_\ast[ O^-, O^-_{\partial L} ] |^2 .
	\end{equation}
	The interior wavefunctional $\Psi^-$ satisfies
	\begin{equation}
		\fl
		\Psi^-_\ast[ O^-, O^-_{\partial L} ] =
		\int [ \d O^-(\tau) ] \; \exp \left( \im S[O^-] \right)
		\int [ \d O^+(\tau) ] \; \exp \left( \im S[ O^+ ] \right) ,
		\label{eq:factorized-path-integral}
	\end{equation}
	where $S$ is a local action functional.
	In this expression,
	$[\d O^-_i(\tau) ]$ represents an integral
	over interior field histories $O^-(\tau, \vect{x})$
	which coincide with
	$O^-(\vect{x})$ at $\tau_\ast$
	and have boundary data $O^-_{\partial L}$ on $\partial L$,
	whereas
	$[ \d O^+(\tau) ]$ represents an otherwise unconstrained
	integral over all exterior
	field histories $O^+(\tau, \vect{x})$ which merge sufficiently smoothly
	with the boundary configuration $O^-_{\partial L}$ on the
	edges of the box.
	
	In typical applications, the expectation
	value~\eref{eq:expectation-value} does not depend on the
	prior history of the universe up to the point of horizon crossing.
	It is therefore independent of almost all boundary data
	represented by $O^-_{\partial L}$, retaining only a possible
	dependence on the boundary configuration near the surface
	$\tau = \tau_\ast$.
	If the path integral is dominated by semiclassical field configurations
	which vary only slowly on the scale $L$, then to a good
	approximation this remaining dependence will involve only the
	roughly homogeneous boundary value at $\tau_\ast$, and not its
	spatial gradients.
	Eq.~\eref{eq:factorized-path-integral} suggests that
	expectation values are to be calculated
	by fixing a homogeneous background field configuration
	at the surface of horizon crossing, and
	summing over unrestricted field histories for fluctuations in the interior.
	Finally, one averages over the fixed background configurations
	with an unknown probability measure, $P$,
	which derives from the integral over
	exterior field configurations.
	At this point we have arrived at the prescription of \S\ref{sec:zoo-box}.

	This construction makes $P$
	a distribution over field configurations at
	a fixed spatial position
	in an ensemble of realizations,
	but if ergodicity applies we can equally interpret
	it as a probability distribution
	over different boxes in the same realization.
	The above argument is merely heuristic, and the existence
	of such a decoupling limit
	has not yet been demonstrated.%
		\footnote{If the probability distribution on the boundary
		does not decouple from spatial
		derivatives, or the field configuration prior to horizon
		crossing, then the mosaic
		approach we have adopted need not apply, and the whole
		question of infrared effects becomes more
		interesting.}
	Nevertheless,
	Eqs.~\eref{eq:expectation-value}--\eref{eq:factorized-path-integral}
	show a
	strong similarity with path integral approaches to
	the operator product expansion (``OPE'')
	in flat space quantum field theory
	\cite{Weinberg:1996kr}.
	A manifestly local and covariant
	formulation of this expansion
	in curved spacetime
	has been given by Hollands \& Wald
	\cite{Hollands:2001nf,Hollands:2001fb,Hollands:2002ux,
	Hollands:2008vx,Hollands:2008ex},
	which in principle could be applied
	to the study of infrared effects in nearly de Sitter spacetimes.
	However, this has not yet been done.
	
	A similar conclusion was reached by
	Allen \& Folacci \cite{Allen:1987tz}
	and later Kumar, Leblond \& Rajaraman \cite{Kumar:2010zx},
	who studied ambiguities in defining the zero-mode
	for the Bunch--Davies
	propagator associated with a massless, minimally
	coupled scalar field in de Sitter space.
	Kumar {\etal}
	noted that similar ambiguities were removed in Minkowski space
	owing to requirements imposed by the principle of cluster
	decomposition. Exploiting the extra freedom
	available in de Sitter space,
	they succeeded in subtracting infrared divergences
	by performing an appropriate redefinition.
	In a field theory this redefinition would naturally
	become spatially dependent on scales which are large compared
	with those appearing in expectation values.
	The result is to
	reproduce the conclusion that infrared divergences derive from
	a supposition that calculations are to be carried
	out with a definite, homogeneous expectation value for the
	background field configuration.
	From this point of view,
	Eq.~\eref{eq:factorized-path-integral} can be thought of
	as an analogue of cluster decomposition in de Sitter.
	
	These ideas have a large literature of their own.
	Kirsten \& Garriga \cite{Kirsten:1993ug}
	used related reasoning to represent the quantized massless scalar
	field as the product of a Hilbert space, associated with the zero-mode,
	and a Fock space corresponding to excitations of finite wavenumber.
	Related ideas were explored by
	Moncrief
	\cite{Moncrief:1976un},
	Higuchi and collaborators
	\cite{Higuchi:1991tk,Higuchi:1991tm,Higuchi:2001uv,Higuchi:2002sc},
	and later
	Giddings \& Marolf \cite{Giddings:2007nu}.
	Urakawa \& Maeda investigated the same physics from
	a different perspective \cite{Urakawa:2007dm,Urakawa:2008rb}.
	
	\subsection{The $L$-dependence of physical predictions}

	While discussing logarithms generated by ultraviolet effects,
	Senatore \& Zaldarriaga remarked that combinations such
	as $\ln \Lambda/k$ in Eq.~\eref{eq:uvlog} were unphysical
	\cite{Senatore:2009cf}, because
	there is no physical comoving scale which could accompany $k$.
	However,
	this argument does not entirely
	preclude the appearance of $k$ in any logarithm.
	For example, the combination $\ln |k\tau_\ast|$
	measures by how many e-folds the wavenumber $k$ is outside the
	horizon at time $\tau_\ast$ and is a sensible physical quantity.
	Nevertheless, this example shows that one must be careful
	in dealing with the combination $\ln kL$, in which $L$ is also a comoving
	scale.
	
	Let us first idealize to an inflating volume which is spatially infinite.
	As in \S\ref{sec:zoo-box}
	we divide this volume into boxes of size $L$,
	in each of which the field is
	spatially homogeneous with small perturbations.
	We have seen that correlations computed within these boxes depend
	on the arbitrary scale $L$, but
	(assuming only zero-modes need be retained from the
	long wavelength physics)
	an answer independent of $L$
	is obtained if we average the correlations over boxes
	\cite{Seery:2009hs}.
	This requires a knowledge of the
	homogeneous background field within each box of the mosaic.
	If merely probabilistic answers are acceptable,
	it may only be necessary to use
	information about the statistical distribution
	of field values over the ensemble of boxes.
	Cosmological perturbation theory has been discussed from this point
	of view by many authors; see, for example,
	the recent textbook by Mukhanov \cite{Mukhanov:2005sc}.

	If we assume that only zero-modes need be retained,
	as in \S\ref{sec:mosaic},
	it follows that
	this argument does not depend on a careful discussion
	of how the total
	$L$ dependence can somehow cancel between the $L$-dependent
	box-sized correlations and the distribution of field values
	among boxes.
	The schematic argument of
	Eqs.~\eref{eq:expectation-value}--\eref{eq:factorized-path-integral}
	shows that in a full treatment, where the entire ultraviolet physics and
	initial conditions are known, we have done nothing more than
	divide the calculation into two parts.
	When we assemble these two halves, the final answer must
	be independent of $L$.
	This happens automatically,
	no matter how remarkably fine-tuned the necessary cancellations
	appear, and
	is a standard argument in the application of effective
	field theories \cite{Burgess:1992gx}.
	The scale $L$ therefore disappears from the answer and we need not
	worry about its physical significance.
	In practice, the distribution of field values among boxes of the
	mosaic will evolve in time. After convolving with this distribution,
	$L$-dependent terms would be traded for a time dependence
	which could itself be represented by secular logarithms.
	
	However, it may not be the case that this $L$-independent answer is
	the quantity we wish to compare with observation.
	In discussing similar questions almost twenty years ago,
	Salopek \& Bond remarked that we only wish to calculate what we can
	observe
	\cite{Salopek:1990re,Salopek:1990jq}.
	What can we observe?
	Only the density fluctuation within our presently observable
	patch of the universe
	\cite{Mollerach:1990zf,
	Lyth:2006gd,Lyth:2007jh,Bartolo:2007ti,Enqvist:2008kt}.
	Observationally it is a matter of perfect indifference to us
	(although, of course, of surpassing theoretical interest)
	whether the distant, unobserved universe contains many regions
	similar to our own, or whether all distant regions are quite
	dissimilar. Whichever is correct, the answer will not change the result
	of satellite observations made today.
	
	If we subscribe to programme outlined in \S\ref{sec:intro},
	then we must suppose that our observable region of the universe
	passed through a phase of inflation which eventually came to an end.
	We assume that the classical background associated with whatever
	scalar fields supplied the necessary vacuum energy
	became roughly homogeneous within
	a box somewhat larger than our present horizon.
	This box should be large enough to
	contain many regions of comparable size to the presently observable
	universe, but need not be exponentially larger
	\cite{Lyth:2006gd,Lyth:2007jh}.
	We wish to compute correlations in the density fluctuation which
	would be recorded by a typical observer whose local patch
	experienced the
	same gross cosmological history.
	For this reason it may be \emph{incorrect} to compare
	the $L$-independent answer discussed above with experiment,
	because in this answer
	the correlations specific to our patch are
	commingled with correlations recorded by observers
	experiencing a gross cosmological history
	quite different to our own.
	
	Therefore, let us choose $L$ so that $\ln kL \sim 1$,
	making $L$ a little larger than our presently observable
	universe.%
		\footnote{This prescription seems to have been rediscovered
		several times, beginning with Salopek \& Bond
		\cite{Salopek:1990jq,Salopek:1990re}.
		In our present context
		the choice $\ln kL \sim 1$ was suggested by Boubekeur \& Lyth
		\cite{Boubekeur:2005fj}, who referred to it as a `minimal box.'
		Its properties
		were later studied in a series of papers by
		Lyth and collaborators
		\cite{Lyth:2005fi,Lyth:2006gd,Lyth:2007jh,Kohri:2009ac}.
		The same prescription, in various forms, was
		given by Bartolo {\etal} \cite{Bartolo:2007ti}
		and later by Enqvist {\etal} \cite{Enqvist:2008kt}
		and Kumar {\etal} \cite{Kumar:2009ge}.}
	The problem we face in this scenario is to determine
	the correct homogeneous background in which we should carry out
	our calculation.
	In general the answer depends on the distribution
	of field values within the ensemble of boxes,
	but there is a simple 
	class of models
	in which this question is trivial.
	In any single-field model with a unique reheating minimum
	there is a unique value of the background field,
	$\phi = \phi_{60}$, when inflation is of order 60 e-folds
	from ending.
	The prior history of the large-scale universe is
	irrelevant
	\cite{Unruh:1998ic,Bartolo:2007ti}.
	To compute what would be recorded by a typical observer in this model,
	we are entitled to carry out
	our calculation within a box
	carrying an approximately homogeneous background field
	with value $\phi \sim \phi_{60}$.
	In this picture, the effective field theories
	we use to study slow-roll inflation are to be understood only as
	a description of the process by which an inflationary trajectory arrives
	in its final reheating minimum, causing inflation to end.
	
	In the distant future our observable region of the universe
	will expand to include modes which are presently unobservable.
	Enqvist, Nurmi, Podolsky \& Rigopoulos remarked
	that an inflationary theorist, calculating millions of years in the
	future, would be obliged to work in a box of slightly larger
	size $L' > L$ \cite{Enqvist:2008kt}.
	The argument of Boubekeur \& Lyth \cite{Boubekeur:2005fj},
	which was extended to two loops
	by Bartolo {\etal} \cite{Bartolo:2007ti},
	shows that physical observables compose correctly under
	the process of averaging $L$-sized boxes to make
	an $L'$-sized box.
	
	\subsection{Stochastic inflation}
	\label{sec:stochastic}
	
	In a scenario which is more general than single-field inflation
	there may be many terminal vacua in which the universe
	can reheat, rather than a unique choice.
	In these models
	there is no preferred background configuration---%
	analogous to the single-field configuration $\phi \sim \phi_{60}$---%
	in which we should carry out our computations.
	Even where a terminal vacuum can be selected in advance,
	it was shown by Byrnes, Choi \& Hall that
	the final fluctuations may depend sensitively on the path
	by which the fields arrive at the minimum
	\cite{Byrnes:2008wi,Byrnes:2008zy} (reviewed in
	Ref.~\cite{Byrnes:2010em}).

	To determine what would be recorded by a typical observer
	we must know at least
	the frequency with which these minima are populated
	by boxes in which inflation terminates.
	This can be ascertained only if we
	know the distribution of field
	values within boxes of the mosaic,
	and (as we have discussed above) this distribution is an
	ultraviolet-dependent quantity. Without knowledge of the complete
	ultraviolet physics, including any relevant initial conditions,
	we cannot calculate it from first principles.

	Apart from our interest in observational predictions,
	there may be other reasons to enquire about the large-scale
	disposition of the scalar fields.
	Senatore \& Zaldarriaga
	\cite{Senatore:2009cf} emphasized that we would like to know
	the circumstances under which eternal inflation can occur
	\cite{Creminelli:2008es,Leblond:2008gg}.
	It is possible this depends on
	effects arising from loops at high order
	in perturbation theory.
	For this reason we may wish to go beyond the narrow view
	advanced above, according to which our theories of slow-roll
	inflation should be
	restricted to a description of how inflation ends.
	Using a model of slow-roll inflation to describe evolution
	within a large spatial volume over very long times would again
	require us to know the distribution of field values over the entire
	mosaic of boxes.
	
	Clearly we cannot hope to calculate this distribution \emph{ab initio}.
	Suppose, however, that by some means we know the distribution
	of field values over the mosaic of boxes at time $t_i$,
	at which point we suppose the Hubble scale took the value $H_i$.
	If we have an effective description of the processes by which
	correlations are established at energies lower than $H_i$,
	then we \emph{can} calculate the subsequent
	evolution of the mosaic.
	
	Similar techniques are used in many applications
	where the long-range physics likewise cannot be determined from
	first principles. A simple example is the distribution of
	partons within colliding nuclei
	\cite{Dissertori:2003pj,Seery:2009hs}.
	Since this distribution depends on the unknown details of
	confinement in QCD it cannot be calculated from first
	principles.
	However, its evolution with energy can be calculated
	at energies large enough that the perturbative regime of
	QCD is a good description.
	In the inflationary case, it is the regime of small
	$H/\Mp$ which is accessible to perturbative calculations.
	A simple method of calculating this evolution was suggested by
	Starobinsky
	\cite{Starobinsky:1986fx}
	and later applied to the calculation of
	correlation functions in the deep infrared
	by Starobinsky \& Yokoyama \cite{Starobinsky:1994bd}.
	
	In Starobinsky's method we suppose that the background field
	value within a given box of the mosaic is known.
	Our discussion of time evolution in~\S\ref{sec:zoo-time}
	and~\S\ref{sec:time} shows that we will obtain reasonable
	answers only if we account for the slow roll-down of this field.
	In a single-field scenario
	the evolution of the background field value is given by
	\begin{equation}
		\dot{\phi} = - \frac{V'}{3H} + \frac{H^{3/2}}{2\pi} \theta(t) ,
		\label{eq:langevin}
	\end{equation}
	where $V(\phi)$ is the potential and
	$\theta(t)$ is a Gaussian noise term satisfying
	\begin{equation}
		\langle \theta(t) \theta(t') \rangle = \delta(t - t') .
	\end{equation}
	In a careful treatment, we would find that
	this noise term depends on what we assume about the physics responsible
	for generating correlations within this box.
	Eq.~\eref{eq:langevin} is a Langevin equation, which can be
	equivalently expressed as an evoution equation for
	the distribution function of
	field values over the mosaic, $f(\phi)$,
	\begin{equation}
		\frac{\partial f}{\partial t} = \frac{1}{3H}
		\frac{\partial (fV')}{\partial \phi} +
		\frac{H^3}{8\pi^2} \frac{\partial^2 f}{\partial \phi^2} .
		\label{eq:fokker}
	\end{equation}
	Various refinements of this equation are possible,
	which were discussed by Salopek \& Bond
	\cite{Salopek:1990re}.
	
	It has been suggested that stochastic inflation may
	represent one possible method of determining the
	distribution of field values over the entire mosaic of boxes
	\cite{Bartolo:2007ti,Enqvist:2008kt,Seery:2009hs}.
	Although this idea is attractive, there are several drawbacks.
	First, Eqs.~\eref{eq:langevin}--\eref{eq:fokker} depend on an initial
	condition for the distribution $f(\phi)$.
	This initial condition is sensitive to
	the details of ultraviolet physics and is no more calculable
	from first principles than is $f(\phi)$ itself.
	This difficulty is usually circumvented by seeking stationary
	solutions of the Fokker--Planck equation, Eq.~\eref{eq:fokker}.
	However, it is not clear under which circumstances this
	is a reasonable choice, or whether the result genuinely approximates
	what would be obtained if we were to retain the complete ultraviolet
	physics.
	
	Second, the evolution equations~\eref{eq:langevin}
	and~\eref{eq:fokker} depend on what we assume about physics
	at high values of the Hubble parameter, up to
	$H \sim \Mp$. In particular, these equations assume that slow-roll
	inflation is a good description throughout this regime.
	This is analogous to the supposition that the Standard Model
	is a good description of physics at energies from the electroweak
	frontier at $\sim 1 \; \mbox{TeV}$ to up the GUT
	scale at $\sim 10^{16} \; \mbox{GeV}$.
	This assumption allows renormalization group evolution
	of the SU(2), U(1)$_Y$ and strong coupling constants
	to predict a unification near the GUT scale in the same
	way that Eq.~\eref{eq:fokker} allows us to predict the behaviour
	of $\phi$ at values of $H$ far above those which gave rise
	to observable density fluctuations.
	In either case it is perfectly reasonable that 
	unguessed physics intervenes
	to spoil the conclusion. If so, the true evolution of
	the background fields might follow laws very different
	to Eqs.~\eref{eq:langevin} and~\eref{eq:fokker}.
	Alternatively, at sufficiently high values of $H$ the concept
	of a homogeneous background field itself may lose meaning.
	In that case the present discussion would be invalidated,
	becoming relevant only at lower energies.

	It was emphasized above that,
	assuming only zero modes need be retained
	from the quasi-classical physics on large
	scales, cancellation of $L$ was automatic after averaging
	over the distribution of field values within the mosaic.
	Since there is no recipe to construct this distribution,
	it must checked individually for each candidate.
	Unfortunately,
	it is non-trivial to verify that the $L$-dependence carried by
	a solution to Eq.~\eref{eq:fokker} would precisely cancel
	that $L$-dependence of correlations computed within a box $L$.
	At the time of writing, this does not appear to have been verified.
	A complete cancellation would depend on retaining sufficient
	information
	about the relevant long-range physics
	in Eqs.~\eref{eq:langevin}--\eref{eq:fokker},
	and it is far from obvious that
	this is the case.
	However, since the same physics which gives rise to the box-cutoff
	logarithms is used to compute the evolution of $\phi$ in
	Eq.~\eref{eq:langevin}, it is
	at least plausible that the $L$ dependence
	of $f$ may have the requisite properties.
	A similar property
	is exhibited by the parton distribution functions
	which describe the physics internal to colliding nuclei
	\cite{Dissertori:2003pj}.
	A first step in this direction was given by
	Kuhnel \& Schwarz \cite{Kuhnel:2010pp}
	who argued that large-scale correlations were strongly suppressed
	by stochastic effects, leading to infrared finite correlation functions.
	
	\subsection{Dynamically generated masses}
	
	In \S\ref{sec:zoo-time} and \S\ref{sec:time} we discussed the role
	of an evolving background field in suppressing correlations at late
	times, and in \S\ref{sec:dynamical} we reviewed the argument of
	Burgess {\etal} \cite{Burgess:2009bs}, according to which
	the secular logarithms could be resummed into a dynamically generated
	mass. This mass drives any correlations to zero in the far future,
	as the field settles in its minimum and loses memory of its past history.
	
	Eq.~\eref{eq:langevin} and its associated Fokker--Planck equation,
	Eq.~\eref{eq:fokker}, account for both fluctuations and time
	evolution in the background field. A fluctuation
	$\theta$ in Eq.~\eref{eq:langevin} can push the field up the potential.
	At later times the term $-V'/3H$ causes it to roll towards
	the minimum.
	Therefore we would not expect statistical properties
	computed
	using the stochastic framework to suffer from the pathological
	failure to decay
	associated with correlations established using Eq.~\eref{eq:massless}.
	This expectation is borne out in explicit calculations.
	
	Starobinsky \& Yokoyama worked with a model of $\lambda \phi^4$
	inflation,
	defined by the interaction of Eq.~\eref{eq:phifour},
	and calculated the evolution of correlations in this theory
	using the stochastic method \cite{Starobinsky:1994bd}.
	Working in the massless case for simplicity, they were able to
	use the Fokker--Planck equation~\eref{eq:fokker}
	to derive an evolution equation for the two-point function
	$\langle \phi^2 \rangle$,
	\begin{equation}
		\frac{\partial}{\partial t} \langle \phi^2 \rangle =
		\frac{H^3}{4\pi^2} - \frac{\lambda}{9H} \langle \phi^2 \rangle^2 .
	\end{equation}
	Starobinsky \& Yokoyama observed that the solution of this
	equation gave a constant,
	\begin{equation}
		\langle \phi^2 \rangle = \frac{3}{2\pi} \frac{H^2}{\sqrt{\lambda}} ,
		\label{eq:sy}
	\end{equation}
	at late times, and remarked that this was equivalent to the
	dynamical generation of a mass of order $H$.
	We briefly review this conclusion using the analysis of
	Riotto \& Sloth \cite{Riotto:2008mv}, who derived a ``gap''
	equation
	for the $\phi$ propagator, $G$,
	\begin{equation}
		\left( \Box + \frac{\lambda}{6}
			\left[ \langle \phi^2 \rangle + G(x,x') \right]
		\right) G(x,x') = \frac{\im}{\sqrt{-g}} \delta( x - x' )
		\label{eq:gap}
	\end{equation}
	where the differential operator $\Box = g^{ab} \nabla_a \nabla_b$
	operates on the variable $x$, and $\nabla_a$ is a covariant derivative
	compatible with the metric $g_{ab}$.
	Eq.~\eref{eq:gap} is obtained using the method of the
	2-particle irreducible action. For details the reader may consult
	the original article by Riotto \& Sloth \cite{Riotto:2008mv},
	or the textbook by Calzetta \& Hu \cite{Calzetta:book}.
	After translating to Fourier space and
	substituting Starobinsky \& Yokoyama's
	late-time solution for $\langle \phi^2 \rangle$,
	given by Eq.~\eref{eq:sy}, in~Eq.~\eref{eq:gap}
	one finds that the the Green's function $G(x,x')$ should be built
	out of solutions to the homogeneous equation
	\begin{equation}
		\ddot{G} + 3 H \dot{G} +
		\left( \frac{k^2}{a^2} + \frac{\sqrt{\lambda}}{4\pi} H^2
		\right) G = 0 ,
		.
	\end{equation}
	This is the Klein--Gordon equation for a field of mass
	\begin{equation}
		\meff^2 = \frac{\sqrt{\lambda}}{4\pi} H^2 .
	\end{equation}
	Later, Burgess {\etal} \cite{Burgess:2009bs}
	argued that this mass could be
	reproduced from the dynamical mass generated by the
	renormalization group, Eq.~\eref{eq:npmass},
	at least in the large-$N$ limit of an $O(N)$-symmetric theory
	similar to that studied by Riotto \& Sloth
	\cite{Riotto:2008mv} and Petri \cite{Petri:2008ig}.

	\subsection{Tensor-to-scalar ratio}
	
	If an exact shift symmetry protects the potential of some field,
	then it experiences no classical evolution.
	An example might be a
	Goldstone boson or axion, or more generally any isocurvature field
	which can be interpreted as an angle. Once perturbations
	have been generated in such a field, they do not decay.%
		\footnote{During inflation,
		quantum fluctuations will completely randomize
		the value of such a field. Nevertheless, the existence of a shift
		symmetry means the field value is unobservable
		and there seem no unacceptable consequences associated
		with this behaviour. If the field is massless but interacting%
		---such as at a critical point---then the field value is
		in principle observable.
		(This example was suggested by C. Burgess
		and L. Leblond.)
		For sufficiently strong interactions the field
		presumably rolls down the resulting potential at late
		times, as described by Senatore \& Zaldarriaga
		\cite{Senatore:2009cf},
		preventing an unphysical divergence to arbitrarily
		large field values. However,
		it is not known
		what conditions must be satisfied by
		the interactions for divergences to be absent.}
	In this case we would not expect resummation to generate a mass,
	and there is no prediction for the value taken by such a field
	within our presently observable patch, although it may be possible
	to estimate a distribution
	\cite{Lyth:2006gd,Kohri:2009ac}.
	Hertzberg, Tegmark \& Wilczek \cite{Hertzberg:2008wr}
	argued that such scenarios represented a test case of anthropic
	reasoning.
	
	What would happen if a microwave background
	observable depended on the value of such a field?
	We would not be able to predict which measurements
	should be recorded by
	CMB survey satellites, such as
	the WMAP and Planck satellite missions.
	A similar argument was given by Sloth, who pointed out
	that an observable such as the tensor fraction, $r$,
	may depend on a ratio of two quantities which receive loop
	corrections scaling differently as the box size increases
	\cite{Sloth:2006az,Sloth:2006nu}.
	The expected value of $r$ will depend on the box in which
	correlations are computed. However, as has been stressed
	throughout, to obtain observables one must compute
	a conditional expectation value,
	contingent on inflation ending in our own vacuum
	(almost surely) everywhere within the box.
	For this reason, the observational
	relevance of infrared corrections in $r$ is not yet clear.
	
	\section{Discussion}
	
	Infrared effects have attracted considerable interest, but at
	the time of writing it has not been possible to extract a
	definitive prediction of a novel effect from these calculations---that
	is, an effect
	which is not already implicit in calculations carried out locally
	around background fields set at a point on the scalar potential.

	The best studied examples of infrared effects
	are the time-dependent secular logarithms
	introduced in \S\ref{sec:zoo-time}.
	Over short timescales these logarithms represent evolution of
	correlations, as the background fields slowly roll down their
	potentials. At later times quantum effects can become important,
	but their role remains somewhat unclear.
	The calculation of Burgess {\etal}, supported by the
	arguments of Senatore \& Zaldarriaga, suggests that in stable theories
	they will suppress correlations at late times.
	
	Box-cutoff logarithms represent a different problem:
	the possibility that, after many e-folds of inflation,
	spatial gradients develop in the background scalar fields.
	If we only wish to use our theories of inflation to study
	observable correlations generated on approach to
	reheating, then this may not be a significant problem.
	On the other hand, if we wish to study the evolution of a large
	spatial volume undergoing inflation for a long period of time,
	we expose ourselves to the possibility of unknown new physics at
	large values of $H/\Mp$. The gauge invariant arguments of
	Losic \& Unruh appear to cast doubt on the idea that perturbation
	theory remains globally well defined for an indefinite period in
	de Sitter or nearly de Sitter spacetimes.

	Except in special cases,
	such as single-field inflation, the study of correlations generated near
	exit from inflation is still ultraviolet-sensitive
	(in the sense of physics at large values of $H/\Mp$)
	and infrared-sensitive
	(in the sense of physics on very large lengthscales),
	because we do not know in which reheating minimum
	inflation is likely to terminate. This need not be a problem
	if observation can accurately determine the properties of our local
	minimum. In that case we would be able to understand
	all the properties of our vacuum, but
	we would not be able to understand how
	to embed this minimum in the larger landscape of the inflationary
	potential. However, this problem is not new and has been with us since
	the early days of the inflationary paradigm.
	
	\ack
	
	My understanding of the ideas discussed in this article is due
	to discussions with my colleagues and collaborators.
	I would like to thank Peter Adshead, Nicola Bartolo,
	Christian Byrnes, Xingang Chen,
	Emanuela Dimastrogiovanni, Richard Easther,
	Louis Leblond, James Lidsey,
	Eugene Lim, Bojan Losic, David Lyth, Karim Malik, Sami Nurmi,
	Dmitry Podolsky,
	Antonio Riotto, Sarah Shandera,
	Martin Sloth, Andrew Tolley, Filippo Vernizzi and Daniel Wesley
	for sharing many insights concerning inflation generally,
	and the treatment of de Sitter loop
	diagrams in particular.
	Clifford Burgess,
	Xingang Chen,
	Richard Holman,
	Louis Leblond, Sarah Shandera and Martin Sloth
	offered helpful criticisms on an early
	typescript of this review.
	I would like to thank Peter Adshead for
	a significant investment of time in
	verifying my recomputations of the inflaton self-loop,
	spectator-loop
	and graviton-loop diagrams discussed in \S\ref{sec:zoo-new}.
		
	\end{fmffile}
	
	\section*{References}
	
\providecommand{\href}[2]{#2}\begingroup\raggedright\endgroup

\end{document}